\documentclass[sigconf]{acmart}
\usepackage{arydshln}
\usepackage{multirow}
\usepackage{threeparttable}
\usepackage[percent]{overpic}
\usepackage{arydshln}
\usepackage{booktabs}
\usepackage{amsmath}
\usepackage{xspace}
\usepackage{wrapfig}
\usepackage[ruled]{algorithm2e}

\DeclareMathOperator*{\argmax}{arg\,max}
\DeclareMathOperator*{\argmin}{arg\,min}
\usepackage{bm}
\usepackage{makecell}

\usepackage{float}
\usepackage{caption}
\usepackage{subcaption}

\AtBeginDocument{%
  \providecommand\BibTeX{{%
    \normalfont B\kern-0.5em{\scshape i\kern-0.25em b}\kern-0.8em\TeX}}}

\setcopyright{acmcopyright}

\copyrightyear{2023} \acmYear{2023} \setcopyright{acmlicensed}\acmConference[SA Conference Papers '23]{SIGGRAPH Asia 2023 Conference Papers}{December 12--15, 2023}{Sydney, NSW, Australia}
\acmBooktitle{SIGGRAPH Asia 2023 Conference Papers (SA Conference Papers '23), December 12--15, 2023, Sydney, NSW, Australia}
\acmPrice{15.00}
\acmDOI{10.1145/3610548.3618205}
\acmISBN{979-8-4007-0315-7/23/12} 

\acmSubmissionID{619}

\newcommand{\name}{\ZY{C$\cdot$ASE}\xspace}

\begin{document}

\title{\name: Learning Conditional Adversarial Skill Embeddings for Physics-based Characters}

\author{Zhiyang Dou}
\email{zhiyang0@connect.hku.hk}
\orcid{0000-0003-0186-8269}
\authornote{Work done during an internship at Tencent AI Lab}
\affiliation{%
  \institution{The University of Hong Kong}
  \country{Hong Kong}
}

\author{Xuelin Chen}
\email{xuelin.chen.3d@gmail.com}
\orcid{0009-0007-0158-9469}
\authornote{Corresponding author}
\affiliation{%
  \institution{Tencent AI Lab}
  \country{China}
}
\author{Qingnan Fan}
\email{fqnchina@gmail.com}
\orcid{0000-0003-1249-2826}
\affiliation{%
  \institution{Tencent AI Lab}
  \country{China}
}

\author{Taku Komura}
\email{taku@cs.hku.hk}
\orcid{0000-0002-2729-5860}
\affiliation{%
  \institution{The University of Hong Kong}
  \country{Hong Kong}
}

\author{Wenping Wang}
\email{wenping@tamu.edu}
\orcid{0000-0002-2284-3952}
\affiliation{%
  \institution{Texas A\&M University}
  \country{USA}
}

\renewcommand{\shortauthors}{Dou et al.}

\newcommand{\ZY}[1]{{\color{black}#1}}
\newcommand{\XL}[1]{{\color{black}XL:#1}}
\newcommand{\QN}[1]{{\color{black}QN:#1}}
\newcommand{\TODO}[1]{{\color{black}TODO:#1}}
\newcommand{\todo}[1]{{\color{black}todo:#1}}
\newcommand{\update}[1]{{\color{black}#1}}
\newcommand{\rv}[1]{{\color{black}#1}}



\begin{teaserfigure}
    \centering
 \vspace{-4mm}
         \includegraphics[width=\textwidth]{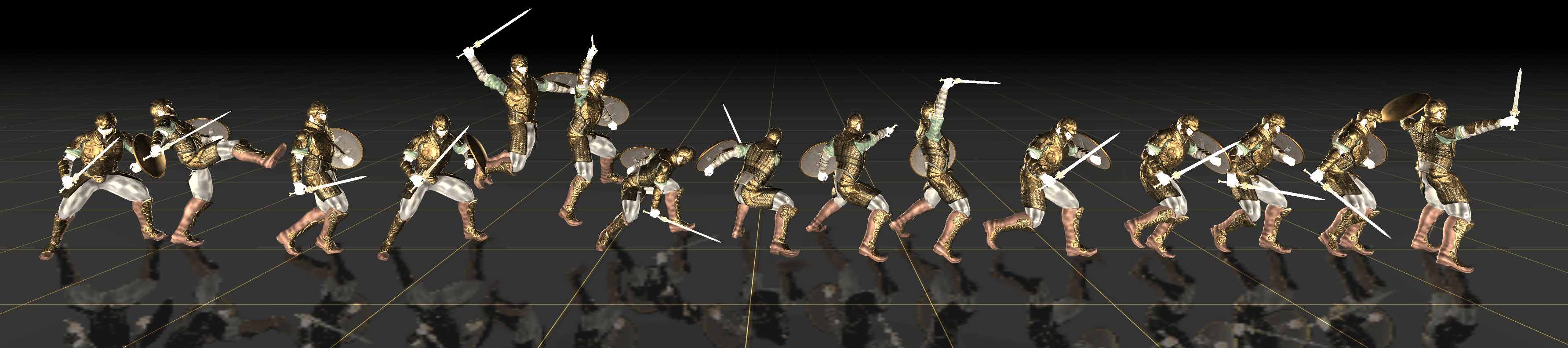} 
         \vspace{-6mm}
  \caption{Our framework enables physically simulated characters to master highly varied and extensive skills with high efficiency and effectiveness.  Notably, it offers an explicit control handle for directly specifying the desired skill from a diverse and extensive set of skills.  Here, a character is instructed by the user to perform a sequence of skills, including kick, jump attack, sword bash, shield bash, and finally, roaring. }
  \Description{This is the teaser figure for the article.}
  \label{fig:teaser}
\end{teaserfigure}
\begin{abstract}
We present \name, an efficient and effective framework that learns Conditional Adversarial Skill Embeddings for physics-based characters. 
\rv{\name enables} the physically simulated character to learn a diverse repertoire of skills while providing controllability in the form of direct manipulation of the skills to be performed.
\rv{This is achieved by dividing}
the heterogeneous skill motions into distinct subsets containing homogeneous samples for training a low-level conditional model to learn the conditional behavior distribution. 
The skill-conditioned imitation learning naturally offers explicit control over the character's skills after training.
The training course incorporates the focal skill sampling, skeletal residual forces, and element-wise feature masking to balance diverse skills of varying complexities, mitigate dynamics mismatch to master agile motions and capture more general behavior characteristics, respectively. 
Once trained, the conditional model can produce highly diverse and realistic skills, outperforming state-of-the-art models, and can be repurposed in various downstream tasks. In particular, the explicit skill control handle allows a high-level policy or a user to direct the character with desired skill specifications, which we demonstrate is advantageous for interactive character animation.
\end{abstract}
\maketitle

\newcommand{\refmotiondataset}{\mathcal{M}}
\newcommand{\motionclip}{m}
\newcommand{\action}{\mathbf{a}}
\newcommand{\skilllabel}{c}
\newcommand{\states}{\mathbf{s}}
\newcommand{\policy}{\pi}
\newcommand{\hpolicy}{\omega}
\newcommand{\goal}{\mathbf{g}}

\newcommand{\skilllatentspace}{\mathcal{Z}}
\newcommand{\skilllatentembedding}{h_z}

\newcommand{\valuefunction}{V}

\newcommand{\skillnum}{K}
\newcommand{\skillset}{\mathcal{C}}
\newcommand{\skilllabelembedding}{h_s}
\newcommand{\skilllatent}{\mathbf{z}}

\newcommand{\likelihood}{d}
\newcommand{\identitymatrix}{\textbf{I}}
\newcommand{\gaussian}{\mathcal{N}}
\newcommand{\prior}{p}

\newcommand{\DKL}{D_{KL}}

\newcommand{\Tsim}{T^{sim}}
\newcommand{\Ttrain}{T^{train}}

\newcommand{\rewardt}{r_t}
\newcommand{\reward}{r}

\newcommand{\imitationreward}{r_I}
\newcommand{\mappingreward}{r_M}
\newcommand{\diversityreward}{r_D}
\newcommand{\mappinglambda}{\lambda_M}
\newcommand{\diversitylambda}{\lambda_D}


\newcommand{\goalreward}{r_G}
\newcommand{\stylereward}{r_S}
\newcommand{\goallambda}{\lambda_G}
\newcommand{\stylelambda}{\lambda_S}

\newcommand{\samplingweights}{w}
\newcommand{\score}{b}
\newcommand{\forgetrate}{\alpha}

\newcommand{\features}{\mathbf{f}}
\newcommand{\maskingmatrix}{\mathbf{T}}
\newcommand{\maskv}{\textbf{h}}
\newcommand{\maskingrate}{\rho}

\newcommand{\encoderq}{q}
\newcommand{\encmu}{\mu}
\newcommand{\encZ}{Z}
\newcommand{\enckappa}{\kappa}

\newcommand{\Discriminator}{D}
\newcommand{\weightGP}{{w}_{\text{gp}}}
\newcommand{\Encoder}{E}

\newcommand{\dataBuffer}{\mathcal{B}}
\newcommand{\maxstep}{E}
\newcommand{\episodelength}{T}
\newcommand{\skilllabelset}{C}
\newcommand{\skilllatentset}{Z}
\newcommand{\trajectory}{\tau}
\newcommand{\updatestepnum}{n}
\newcommand{\numtransitions}{Q}
\newcommand{\goalposition}{\mathbf{x}_g}
\newcommand{\highdiscovery}{L^E}
\newcommand{\highdiscretediscovery}{L^E_c}
\newcommand{\highcontinuousdiscovery}{L^E_z}
\newcommand{\highcontinuousweights}{\lambda^h_z}

\newcommand{\filteringrate}{{filtering rate}}
\newcommand{\filteringrates}{{filtering rates}}

\newcommand{\filteringrategamma}{\gamma}

\newcommand{\framenum}{L}

\section{Introduction}
Humans possess a remarkable ability to acquire a wide range of skills through years of practice and can effectively utilize these skills to handle complex tasks. 
Generally, the motion of these diverse skills is \textit{heterogeneous}. 
For instance, stationary standing and highly dynamic back-flipping exhibit significant differences in terms of their movement over time. In practice, humans can divide these heterogeneous skill motions into homogeneous sets so as to focus on a specific skill during each training session.
This divide-and-conquer idea is classic and has demonstrated efficacy in the field of physics-based character control~\cite{won2020scalable, liu2018learning}.

However, this is at odds with \rv{recent} state-of-the-art literature in \ZY{learning a large-scale, reusable and unified embedding space for} physics-based characters, where samples (i.e., state transitions) generated from diverse skills with heterogeneity are treated as \textit{homogeneous} and learned collectively into latent representations~\cite{ase, controlvae, cvae_meta}.
Apparently, transitions of an idle motion differ significantly from those in a sword-swinging motion. 
This inconsistency undermines the performance of existing models that assume learning on homogeneous samples, as evidenced by the severe \textit{mode collapse}~\cite{ase, controlvae}. 
Moreover, these holistically embedded skill representations do not provide controllability in the form of direct specification of the desired skills, which is a highly desirable property of character animation for offering an interactive and immersive experience to the user.
Although some have shown skill-level control with additional training to stimulate corresponding emergent behaviors from holistically embedded skills~\cite{controlvae}, they have not demonstrated the scalability to extensive skills.

In this work, we advocate that the heterogeneous nature of diverse skills, combined with the need for controllability over embedded skills, necessitates a novel training paradigm for efficient skill learning. To this end, \ZY{we employ the classic divide-and-conquer strategy and present \name,} an efficient and effective framework that learns \text{C}onditional \text{A}dversarial \text{S}kill \text{E}mbeddings for physics-based characters. While learning individual skills had been adopted in training physics-based character~\cite{won2020scalable, liu2018learning},
a particular emphasis of this work is to scale it up to large-scale motion datasets.
Specifically, given a large dataset containing diverse motion clips and labels, which can be annotated per clip manually or using a neural skill labeler (see Appendix~\ref{ap:skill_acquisition}), a low-level latent-variable model is conditioned on skill labels to capture the conditional behavior distribution, and is trained via conditional adversarial imitation learning. The conditional model can learn extensive skills efficiently while also naturally offering explicit control over the character's skills.

That being said, we have to address unique challenges arising from training our characters. First, efficient training should distribute the learning resources, such as training time, over skills with different levels of complexity. For instance, imitating energetic sword swinging is apparently harder than idling. This is achieved by the \emph{focal sampling strategy} in our framework, which encourages the training to dynamically mine hard skill samples and eventually leads to faster and balanced coverage of diverse reference skills. In addition, incorporating a vast array of heterogeneous skills leads to challenges or even failure in adversarial imitation learning, which has also been revealed and explained by the dynamics mismatch between the virtual character and real human in~\cite{yuan2020residual}. 
Hence, we resort to the \emph{skeletal residual forces} to augment the character's control policy when learning highly varied motions, improving the motion quality of particularly agile skills. 
Last, as samples under each skill are sparse,
we further adopt an \emph{element-wise feature masking}, which is simply realized by introducing dropout layers inside the discriminator to avoid over-reliance on motion details and hence enable capturing general behavioral characteristics, leading to diversified transitions learned under each skill condition.

Once pre-trained, our conditional model can produce highly diverse and realistic skills and offers an explicit control handle for direct manipulation of the character's skills. Experiments show our model achieves state-of-the-art performance in capturing the reference motion distribution, outperforming competing methods by a significant margin (coverage: Ours \-- 91\% vs. CALM~\cite{calm} \-- 71\% vs. ASE~\cite{ase} \-- 66\%). Furthermore, we demonstrate that our conditional model can benefit the character animation by integrating it into an interactive authoring system by training deep RL-based high-level policies, that allows users to manipulate physically simulated characters with discrete skill label specifications and other control signals. This is similar to those in video games, but our model supports a much wider range of skills and produces physically plausible motions. 
Last, we showcase the use of our conditional model in various traditional high-level tasks, where policies learn to direct the low-level conditional model for completing different tasks~(Appendix~\ref{ap:high_level_no_user}).

\section{Related Work}
With advancements in Deep Reinforcement Learning~(DRL) and the accessibility of high-quality motion capture (mocap) datasets~\cite{CMU, SFU, mahmood2019amass, tsuchida2019aist, wang2020unicon, harvey2020lafan}, data-driven methods have demonstrated impressive results in physics-based character animation. In the following, we mainly cover these data-driven methods that fall into two categories: 

\paragraph{Tracking-based Methods} \citet{peng2018deepmimic,bergamin2019drecon,park2019learning,won2020scalable, fussell2021supertrack} train controllers to imitate reference motions by tracking target pose sequences from motion clips. DeepMimic~\cite{peng2018deepmimic}, the pioneering work, trains a policy network with random state initialization and early termination for mimicking. The idea was later extended to track motion matching-generated reference motions for responsive characters~\cite{bergamin2019drecon} and improved by removing motion matching dependency~\cite{park2019learning} with a recurrent neural network predicting future reference poses. Efforts have been made to control the diverse behaviors of physically simulated characters.~\citet{won2020scalable} constructs a large motion graph from mocap data, groups graph nodes into clusters, and trains a mixture of expert networks for tracking. SuperTrack~\cite{fussell2021supertrack} introduces a world model represented by a neural network trained to approximate physical simulation, enabling supervised policy network training and accelerating the process.

However, tracking-based methods typically struggle to imitate various skills from large, diverse motion datasets. Composing disparate skills often requires a dedicated motion planner to select appropriate clips for complex tasks, which eludes these methods.

\paragraph{Learning Skill Priors}
The prevailing trend in physics-based character is to learn powerful skill priors from large and diverse motion datasets. 
By embedding distinct skills into a low-dimensional latent space, these models can reproduce versatile skills and be reused to learn high-level controllers for complex tasks. \citet{merel2018neural} distill skill expert networks into a latent space for high-level tasks, while Catch and Carry~\cite{merel2020catchandcarry} incorporates vision signals for diverse full-body tasks. \ZY{\citet{won2020scalable} utilize a mixture of experts to learn skills from a diverse set of behaviors, where the expert network for each skill needs to be trained individually, and subsequently, a gating network is trained to combine the experts. However, training a large number of experts can be laborious and costly, and learning the gating network to effectively integrate them can be challenging.} \citet{peng2019mcp} propose a hierarchical controller with multiplicative compositional policies for more composable control. Recently, \citet{cvae_meta} employ a conditional variational auto-encoder (VAE) for embedding skills into a low-dimensional Gaussian distribution. \citet{controlvae} further introduces ControlVAE to learn a state-conditioned motion prior. 

In the GAIL regime~\cite{ho2016generative}, \citet{peng2021amp} present Adversarial Motion Prior~(AMP) for goal-conditioned reinforcement learning with life-like motions. However, AMP's high-level objective is coupled with a low-level style reward, necessitating tedious fine-tuning for complex tasks. Later, \citet{ase} introduce Adversarial Skill Embeddings~(ASE) to learn reusable skill priors, with which high-level controllers learn to direct for complex tasks. \cite{juravsky2022padl} focuses on aligning skill embeddings with a pre-trained language latent space~\cite{clip}, allowing natural language-directed characters.
CALM, a concurrent work that shares a similar setting to ours, introduces a framework for learning semantic motion representations from a large dataset and also demonstrates control over the skills with \emph{known} semantic labels of reference motion clips. Their key is to heuristically align overlapping samples extracted from the same clip while pulling apart those from different clips. Hence, their success is still contingent upon well-semantically segmented clips. As a consequence, CALM's training is unstable and suffers from severe mode collapse, as recognized by the authors and evidenced in our comparisons.

In general, while these methods can generate various motions, they often suffer from mode collapse when handling highly varied and extensive skills. 
We attribute this issue to treating samples from distinct skills as homogeneous ones, which goes against the nature of human motions, and propose decomposing the whole repertoire into homogeneous subsets for learning skill-conditioned behavior distributions with several crucial training techniques. Last, our low-level conditional model provides effective controllability over the embedded skills, allowing a high-level policy or user to direct characters to perform desired skills.

\section{Method}
\begin{figure}[!t]
\centering
\begin{overpic}
[width=0.48\textwidth]{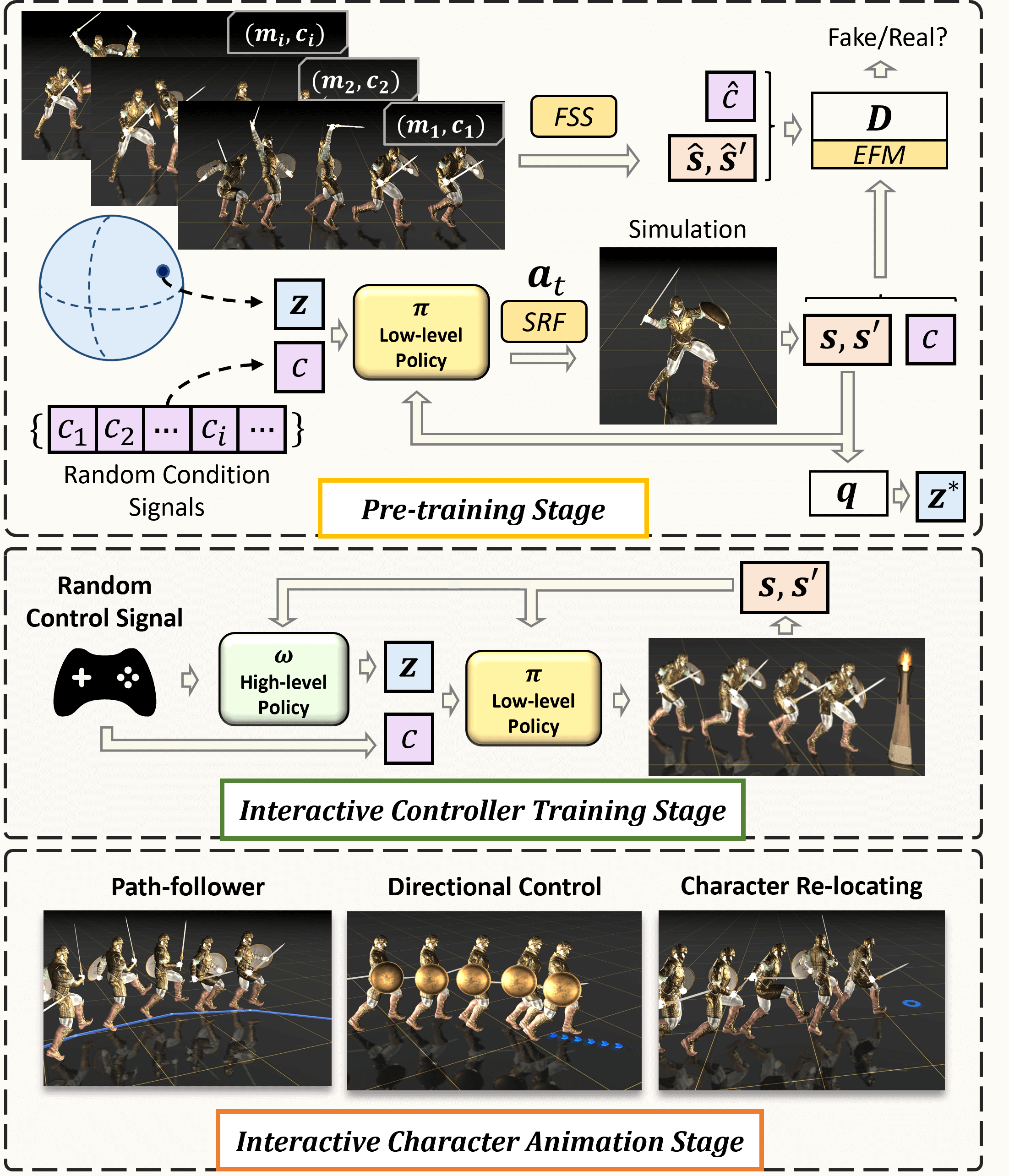}
\end{overpic}
\caption{Our framework contains three stages: the pre-training, interactive controller training, and interactive character animation stages. 
During pre-training,  a low-level policy $\policy$ learns conditional adversarial skill embeddings from a diverse and extensive motion dataset,
followed by more high-level policies $\policy$ trained to allow interactive control of the character. 
Last, during the interactive character animation stage, users can interactively animate the character in various manners, possibly with desired skills.}
\label{fig:pipeline}
\vspace{-5mm}
\end{figure}

Our framework consists of three stages (See Figure~\ref{fig:pipeline}):
1) Pre-training stage, where a low-level conditional policy is trained to imitate reference skills;
2) Interactive controller training stage, where more high-level policies are trained to allow more interactive controls of the character; and 3) Interactive character animation stage, where users can interactively animate the character in various ways.

During the pre-training stage, a reference dataset $\refmotiondataset = \{ (\motionclip^i, \skilllabel^i) \}$ with annotated motion clips $\motionclip^i$ and skill labels $\skilllabel^i$ is used for learning conditional adversarial skill embeddings. 
Each motion clip $\motionclip^i = \{ \states^i_t \}$ is represented as a sequence of states that depicts a particular skill. 
Note that different $\motionclip^i$ can correspond to an identical skill label $\skilllabel^i$ in the dataset. 
Then, a low-level conditional policy $\policy(\action|\states,\skilllatent,\skilllabel)$\footnote{For simplicity, we ignore the superscript and subscript from now on} is trained through conditional adversarial imitation learning, mapping latent variables $\skilllatent$ to behaviors resembling motions specified by $\skilllabel$. 
At the interactive controller training stage, we train additional policies to attain more controls for interactive character animation, such as directional control, 
 path-follower, etc. At the interactive animation stage, trained policies are fixed, and then users can interactively animate the physics-based character by specifying the desired skills, moving directions/paths or target location.

\subsection{Learning Conditional Adversarial Skill Embeddings}
\label{sec:CAIL}

\name divides the dataset into sub-sets that each contain homogeneous samples, 
from which a conditional low-level policy learns a conditional action distribution. We assume that a transition under a skill category $\skilllabel$ is represented by a latent variable $\skilllatent$ sampled from a prior hypersphere distribution $\skilllatentspace$, 
i.e., ${\skilllatent} = \overline{\skilllatent} / \Vert\overline{\skilllatent}\Vert, \overline{\skilllatent} \sim \gaussian(0,\mathit{\identitymatrix})$. Specifically, a low-level policy $\policy(\action|\states,\skilllatent,\skilllabel)$ takes as input the character's current state $\states$, a latent variable $\skilllatent$, and, more importantly, a skill label $\skilllabel$, and then learns to output an action $\action$ that eventually leads to motions conforming to behavioral characteristics specified by motions sampled from the skill indicated by $\skilllabel$.

Building upon the success in~\cite{ase}, a pioneer work learns large-scale adversarial skill embeddings in a GAN-like framework, we train the low-level policy network with a conditional adversarial imitation learning procedure, where the low-level policy $\policy$ learns to fool a discriminator $\Discriminator(\states,\states',\skilllabel)$, 
that is also conditioned on skill labels $\skilllabel$
and learns to distinguish reference motions from generated ones. We train with the imitation objective and diversity objective to capture the conditional behavior distribution implicitly. Specifically, the conditional discriminator is trained to minimize: 
\begin{equation}
\begin{aligned}
    \min_\Discriminator = &-\mathbb{E}_{\likelihood^{\refmotiondataset}(\states,\states', \skilllabel)}\left[\log(\Discriminator(\states,\states', \skilllabel)\right]  \\ &
    - \mathbb{E}_{\likelihood^{\pi}(\states,\states', \skilllabel)} \left[\log(1-\Discriminator(\states,\states',\skilllabel)\right]  \\ &
    +\weightGP\mathbb{E}_{\likelihood^{\refmotiondataset}(\states,\states', \skilllabel)}\left[\Vert\nabla_\phi \Discriminator(\phi)|_{\phi \in (\states,\states',\skilllabel)}\Vert^2 \right],
    \label{eq:imitation_goal}
\end{aligned}
\end{equation}
where $\likelihood_{\refmotiondataset}(\states, \states', \skilllabel)$ and
$\likelihood_{\policy}(\states, \states', \skilllabel)$ denote state
transitions $(\states,\states')$ drawn from the reference skill $\skilllabel$ and ones generated by the conditional policy $\policy$, respectively.
The last term is a gradient penalty regularization for stabilizing the training.


%
To facilitate training $\policy$, we first employ a conditional motion encoder $\encoderq$ to enforce the mapping between state transitions $(\states, \states')$ and the latent $\skilllatent$ under the $\skilllabel$ label-conditioned distribution.
Since the conditional latent space under $\skilllabel$ is modeled as a hypersphere, $\encoderq$ is modeled as a von Mises-Fisher distribution:
$
   \encoderq(\skilllatent|\states,\states', \skilllabel) = \frac{1}{\encZ} e^{\enckappa, \encmu_\encoderq(\states,\states',\skilllabel)^T \skilllatent}, 
$
where $\encmu_{\encoderq}(\states,\states',\skilllabel)$ is the mean of the distribution, which is further normalized by requiring $|| 
\encmu_\encoderq(\states,\states',\skilllabel) ||=1$. 
$\encZ$ and $\enckappa$ represent a normalization constant and a scaling factor, respectively. The encoder network $\encoderq$ for each skill condition $\skilllabel$ is then trained by maximizing the log-likelihood:
\begin{equation}
 \max_{\encoderq} \mathbb{E}_{\prior(\skilllatent)}
  \mathbb{E}_{\likelihood^{\policy}(\states, \states',\skilllabel|\skilllatent)}\left[\enckappa \encmu_{\encoderq}(\states,\states', \skilllabel)^T\skilllatent\right]
\end{equation}

Then, the reward for policy $\policy$ at each simulation time step $t$ is given by:
$
    \rewardt = -\log(1-\Discriminator(\states_t, \states_{t+1}, \skilllabel)) + \beta \log \encoderq(\skilllatent_t|\states_t,\states_{t+1}),
$ where $\beta$ is a balancing factor. We further add the diversity term to the total objective of $\policy$ in a buffer of $\Tsim$ simulation time steps:
\begin{equation}
\begin{aligned}
  &\argmax_\policy = \mathbb{E}_{p(\skilllatentset)}\mathbb{E}_{p(\trajectory|\policy, \skilllatentset, \skilllabelset)}\biggr[\sum_{t=0}^{\Tsim-1}\gamma^t( \rewardt ) \biggr]\\
 &- \diversitylambda  \mathbb{E}_{\likelihood^\policy(\states)} \mathbb{E}_{\skilllatent_1, \skilllatent_2 \sim\prior(\skilllatent)}\left[(\frac{\DKL(\policy(\cdot|\states,\skilllatent_1, \skilllabel), \policy(\cdot|\states,\skilllatent_2, \skilllabel))}{0.5(1-\skilllatent_1\skilllatent_2)}-1)^2\right]
\end{aligned}
\end{equation}

The above description provides a conceptually valid formulation for learning the conditional distribution.  We further elaborate on algorithmic designs that are crucial for training the low-level policy in the following sections.

\paragraph{Focal Skill Sampling (FSS)}
\label{sec:focal_sampling}
Large motion datasets exhibit a notable characteristic wherein the skills contained within vary in terms of their difficulty to learn. This would inevitably cause unbalanced development, i.e., missing a considerable number of skills, if all skills were treated equally. 
\ZY{To overcome this issue}, we devise a focal skill sampling strategy that naturally fits into the conditional imitation learning course, improving the training via adaptively adjusting the sampling across different reference skills. 

At each training step, let $\overline w_{\skilllabel}$ denote the sampling probability of skill $\skilllabel$ into the reference motion buffer, and $\score_c \in \left(0,1\right)$ represent the average score (the probability that the sample is real, i.e., from the reference data) output by the conditional discriminator $\Discriminator$ on the generated motions under skill category $\skilllabel$. The sampling probability is then updated online at each training step as follows:
\begin{equation}
    \samplingweights_{\skilllabel}  = (1-\forgetrate) \samplingweights_{\skilllabel} + \forgetrate \sigma(\score_{\skilllabel}) ,
    \quad
    \sigma(\score_{\skilllabel}) = 1 - {\score_{\skilllabel}} / \textstyle\sum_{\skilllabel \in \skillset} {\score_\skilllabel} ,
\end{equation}

where $\skillset$ denotes the set of skill category labels, $\forgetrate$ control the update rate, and $\samplingweights_{\skilllabel}$ is the evenly initialized sampling weights. Note that we normalize $\overline\samplingweights_{\skilllabel} = \frac{\samplingweights_{\skilllabel}}{\sum_{\skilllabel \in \skillset}\samplingweights_{\skilllabel}}$ to serve as the final sampling probability. Empirically, this strategy is applied after approximately 2500 training steps, allowing $\Discriminator$ to gain discrimination ability first. We demonstrate that this approach enables the policy to effectively cover more skills with high efficiency.

\paragraph{Skeletal Residual Forces~(SRF)}
To augment the control policy to effectively imitate complex and agile motions, such as the jump sidekick, we apply a \ZY{torque} computed from PD target control signals to each joint and require the policy network to predict the residual force at each joint position, effectively compensating for the dynamics mismatch between the virtual character and the real human. \ZY{Note that such skeletal residual forces are applied to the character in both training and inference stages.} The equation of motion for multi-body systems with residual forces is given by:
\begin{equation}
    B(\bm{q})\bm{\ddot q} + C(\bm{q},\bm{\dot q}) + g(\bm{q}) =
    \begin{bmatrix}
    \bm{0} \\ \bm{\tau}
    \end{bmatrix}
    +\underbrace{\sum_i\bm{J}^T_{\bm{v}_i}\bm{h}_i}_{\text{Contact Forces}} + \underbrace{\sum^{J-2}_{j=1}\bm{J}^T_{\bm{e}_j}\bm{\xi}_j}_{\text{Residual Forces}}
\end{equation}
On the left-hand side, $\bm{q}$,  $\bm{\dot q}$, $\bm{\ddot q}$, $B$, $C$, and $g$ represent degrees of freedom (DoFs) of joints, joint velocities, joint accelerations, the inertial matrix, the \ZY{vector} of Coriolis and centrifugal terms, and the gravity vector, respectively. On the right-hand side, the first term comprises the torques  $\bm{\tau}$ computed from the PD target control signals applied to the non-root joint DoFs, whereas $\bm{0}$ denotes the non-actuated root DoFs. The second term describes the contact forces $\bm{h}_i$ on the humanoid (typically exerted by the ground plane) and the contact points $\bm{v}_i$ of $\bm{h}_i$, ascertained by the simulation environment. The Jacobian matrix $\bm{J}_{\bm{v}_i} = d\bm{v}_i/d\bm{q}$ delineates the manner in which the contact point $\bm{v}_i$ varies with respect to the joint DoFs $\bm{q}$. \ZY{The Jacobian matrix $\bm{J}_{\bm{e}_j} = d\bm{e}_j/d_{\bm{q}}$ defines the change of the contact point $\bm{e}_i$ of the residual force $\bm{\xi}_j$ w.r.t. the joint DoFs $\bm{q}$.} In the last term, similar to~\cite{yuan2020residual}, we ask the policy network to predict the residual forces $\bm{\xi}_j$ at contact point $\bm{e}_j$. We set the contact points $\bm{e}_j$ to be $J-2$ body joints $\bm{j}$, excluding the sword and shield. In this paper, the joint number $J$ of the avatar is $17$. We apply regularization to the residual forces, ensuring that the policy employs these forces when necessarily required: $r_{f} = \text{exp}(-\sum^{J-2}_j\bm{\xi}_{j=1})$.

\paragraph{Element-wise Feature Masking (EFM)}
Learning conditional distributions exacerbates data scarcity under each skill, leading to overfitting and reduced motion diversity. Hence, we further adopt an element-wise feature masking strategy, which is simply realized by introducing dropout layers inside the discriminator, to
significantly alleviate this issue. This stochastic operation avoids over-reliance on motion details, enabling the capture of general behavioral characteristics from sparse samples and resulting in diversified transitions under each skill.

\subsection{Interactive Character Animation}
\label{sec:high_level_control}
The learned low-level conditional model can capture extensive and diverse skills and provides explicit control over the character skills. 
Moreover, we train additional deepRL-based high-level policies to support interactively animation of the character in various ways.

\begin{figure*}[t!]
\centering
\includegraphics[width=\linewidth]{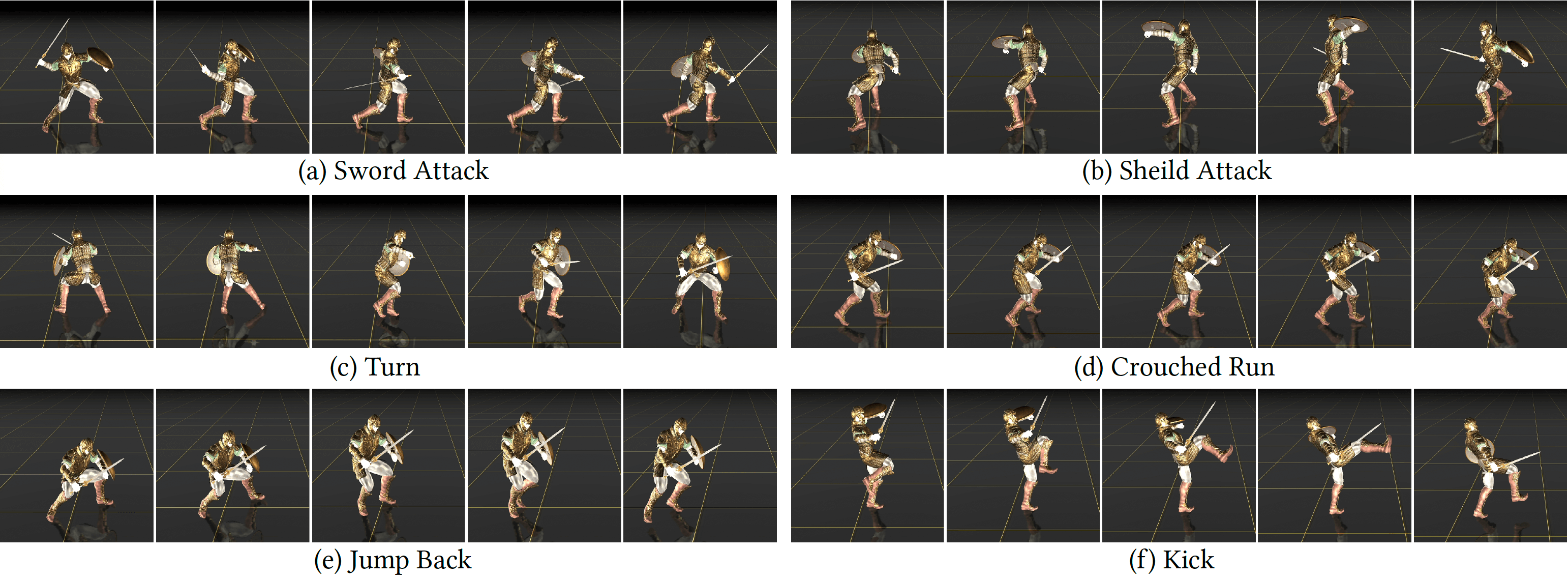}
\caption{\name enables the physically simulated character to perform skills specified by skill labels and transition latent codes.}
\label{fig:low_level_motion}
\end{figure*}


\paragraph{Directional Control.} 
To enable directional control, a high-level policy takes as input the control signals $(\skilllabel_t, d^*_t, h^*_t)$ where $\skilllabel_t, d^*_t, h^*_t$ represents the user desired skill, target local facing direction of the root and the target moving direction, respectively; $^*$ stands for the character's local coordinate frame. The objective of the interactively directional control is given by
\begin{equation}
    r^D_t = 0.7\exp \left(-0.25\Vert\mathbf{v}^*_t - \mathbf{d}^*_t\cdot {\dot{\mathbf{x}}}^{\text{root}}_{t}\Vert^2 \right) + 0.3 \mathbf{h}^*\cdot \mathbf{h}_t^\text{root},
\end{equation}
where $\mathbf{h}_t^\text{root}$ and $\dot{\mathbf{x}}^\text{root}_t$ represent the heading direction and velocity of the character root under specified skill label $\skilllabel_t$. We vary the desired velocity $v^* \in [0,5]$ m/s during training for velocity control. Instead of training the interactive controller to learn to switch skill labels, our low-level model allows for skill switching by explicitly assigning $\skilllabel_t$ at the $t$-th time step, and the interactive controller thus only needs to predict a configuration of latent codes $\skilllatent$ to complete the task under skill $\skilllabel$. During training, we randomize the skill label every five execution steps to simulate user interactive control. After training, users can interactively control the character by specifying the desired skills while dynamically controlling the moving direction akin to those in video games.

\paragraph{Target Location Control} 
We also support re-locating the character to a target location. 
The inputs to the high-level policy are $(\skilllabel_t, x^*_t)$, where $\skilllabel_t, x^*_t$ represents the user desired skill and target location in the character's local frame, respectively. The objective is given by
\begin{equation}
    r^L_t = -0.5\Vert\mathbf{x}^*_t - \mathbf{x}^{\text{root}}_{t}\Vert^2.
\end{equation}
Here, $\mathbf{x}^\text{root}_t$ is the character root location. The high-level controller predicts latent codes $\skilllatent$ to navigate the character through the low-level controller. Skill labels are randomized every five execution steps. 
Once trained, users can dynamically specify a target location and a desired skill label to re-locate the character.

In addition to these interactive controllers,
we also evaluate \name in various representative high-level tasks, such as \textit{Reach}, \textit{Steering}, \textit{Location}, and \textit{Strike} as in~\cite{ase}. 
In these tasks, no user-specified labels are used, and the high-level policies learn to predict the configuration of the skill label and skill latent code for completing the task. For details, please refer to Appendix~\ref{ap:high_level_no_user}.

\section{Experiments}
\label{sec:exp}
We evaluate the efficacy of our framework by training skilled-conditioned control policies for a 3D simulated humanoid character.
Please refer to the supplementary video for more qualitative results.

\paragraph{Dataset.}
We conduct evaluations on two datasets, including: 
1) \emph{Sword\&Shield} dataset from~\cite{ase} containing 87 clips~\footnote{Due to permission issues, the released dataset contains only 87 clips instead of the 187 described in~\cite{ase}, as confirmed by the authors.} and each with a corresponding skill label; 
2) \emph{Composite Skills} dataset, that has $265$ types of skills, including 87 motion clips from~\cite{ase} and 691 clips of 178 manually annotated skills from the CMU Mocap dataset~\cite{CMU}. 
The character, equipped with a sword and shield, has 37 degrees of freedom. 
We retarget the motions from the CMU Mocap dataset to an avatar with a sword and a shield; more details are in Appendix~\ref{ap:training_details}. 
%
Unless specified, we conduct experiments on the Sword\&Shield dataset for a fair comparison with baselines. 
In addition, we also demonstrate the scalability of our method on the Composite Skills dataset in Appendix~\ref{ap:scalability}.

\paragraph{Training}

We train the character in IsaacGym~\cite{isaacgym} with a simulation frequency of $120$ Hz and policy frequency of $30$Hz.
The policies, value functions, encoder and discriminator, are modeled using separate multi-layer perceptions, and the policy networks $\policy$ and $\hpolicy$ are trained with the proximal policy optimization~\cite{ppo}. Policies are trained on a single A100 GPU, with about 1.5 billion samples, corresponding to approximately 1.5 years of simulated time, taking 1.5 days, and high-level policies taking one day. The final animation is retargeted to a rigged avatar. Curriculum learning and joint masking are employed during training; See more details in Appendix~\ref{ap:training_details}.

\subsection{Low-level Conditional Policy}
We first train the low-level policy alone to evaluate its ability to reproduce skills in the motion dataset, particularly when directed by a specified skill label. 
The policy is able to follow a \emph{random} skill label presented to it, such as left sword swing, right shield bash, etc. 
Although the Sword\&Shield dataset contains only one clip of each skill label, the policy is able to perform corresponding skills with \emph{local variations}.  
Examples of behaviors produced by the policy when given various skill labels are shown in Figure~\ref{fig:low_level_motion}. 
Next, we present more quantitative evaluations of the low-level policy. More results are presented in the supplementary video.

\paragraph{\rv{Filtered Motion Coverage Rate}}
\label{sec:motion_coverage_rate}
We evaluate our model in reproducing various motions in the dataset when given random skill labels. Moreover, we compare our model to SOTA methods \--- ASE and CALM, which also train a low-level policy to reproduce skills in the dataset. Note CALM also trains a low-level conditional policy.
All models are trained with the Sword\&Shield dataset released by~\cite {ase, calm}. While there are other important prior works~\cite{juravsky2022padl}
, we were not able to compare exhaustively with them as they have not released the source code. 
Specifically, the quantitative comparison is conducted with the metric motion coverage rate, following~\cite{ase}. 
The trajectories of our model are generated using random skill labels and latent codes, whereas the trajectories of ASE and CALM are obtained with random skill latent codes. 
Furthermore, we propose to compute \emph{filtered} motion coverage to factor out stochastic factors built upon the motion coverage rate proposed by~\cite{ase}.
\label{ap:filtered_motion_coverage}
\ZY{
For each state transition $ (\hat \states_t,\hat \states_{t+1})$ produced from the policy $\policy$ with a skill label $\skilllabel$ and a latent code $\skilllatent$, we find the closest motion clip $m^*$ in the reference motion dataset $\refmotiondataset = \{ (\motionclip^i, \skilllabel^i) \}$:}

\begin{equation}
\motionclip^* = \argmin_{\motionclip^i \in \refmotiondataset} \min_{(\states_t, \states_{t+1})\in \motionclip_i} ||\hat \states_t - \states_t||_2+||\hat \states_{t+1} - \states_{t+1}||_2.
\label{eq:coverage_rate}
\end{equation}
\ZY{
We repeat it for every transition in randomly generated trajectories, and the reference motion clip that contains the best-matched transition will be marked as the one that best matches the trajectory. 

Let $l_i$ denote the number of matched motions under the $i$-th skill category. The \emph{expected} number of samples under each category is given by $\frac{N}{\skillnum}$, where N is the number of generated trajectories and K is the number of skill categories. 
Then, a reference motion skill is identified as covered only if $l_i > \filteringrategamma \frac{N}{\skillnum}$, where $\filteringrategamma$ denotes the filtering rate. Then, the filtered motion coverage rate is given by: }
\begin{equation}
\text{coverage}(\refmotiondataset, \policy, \filteringrategamma) = \frac{1}{\skillnum} \mathcal{I}_{{i\in\{1,\cdots,\skillnum\}}}(l_i > \filteringrategamma \frac{N}{\skillnum}).
\label{eq:filter_coverage_rate}
\end{equation}

\begin{figure}[t!]
\centering
\begin{overpic}[width=0.9\linewidth]{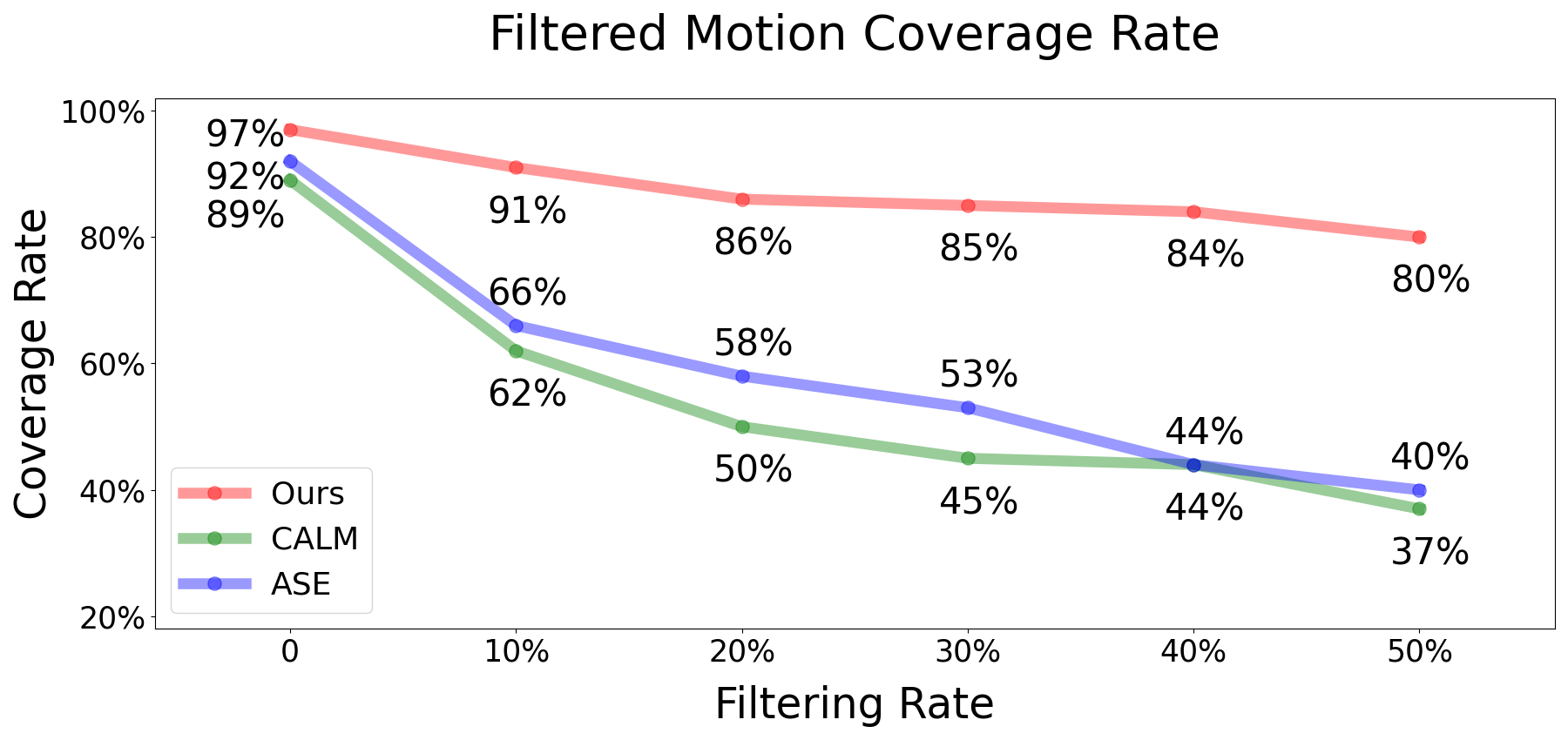}
\end{overpic}
\vspace{-2mm}
\caption{Comparison of the motion coverage. The coverage rate of ASE and CALM falls dramatically with an increasing filtering rate, implying a serious imbalance of the coverage, whereas ours consistently produces high coverage rates.}
\label{fig:motion_covrage_cutoff_curve}
\end{figure}

\begin{figure}[t!]
\centering
\begin{overpic}[width=\linewidth]{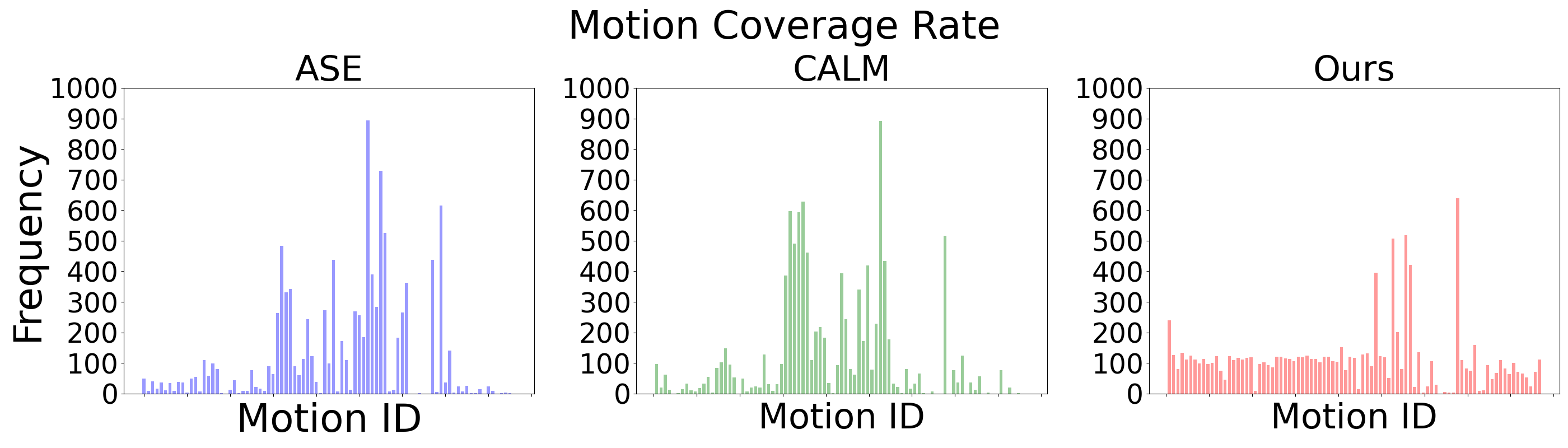}
\put(41,-3){(a) No filtering.}
\end{overpic}
  \makebox[\linewidth][c]{}\\
  \makebox[\linewidth][c]{}\\
  \begin{overpic}[width=\linewidth]{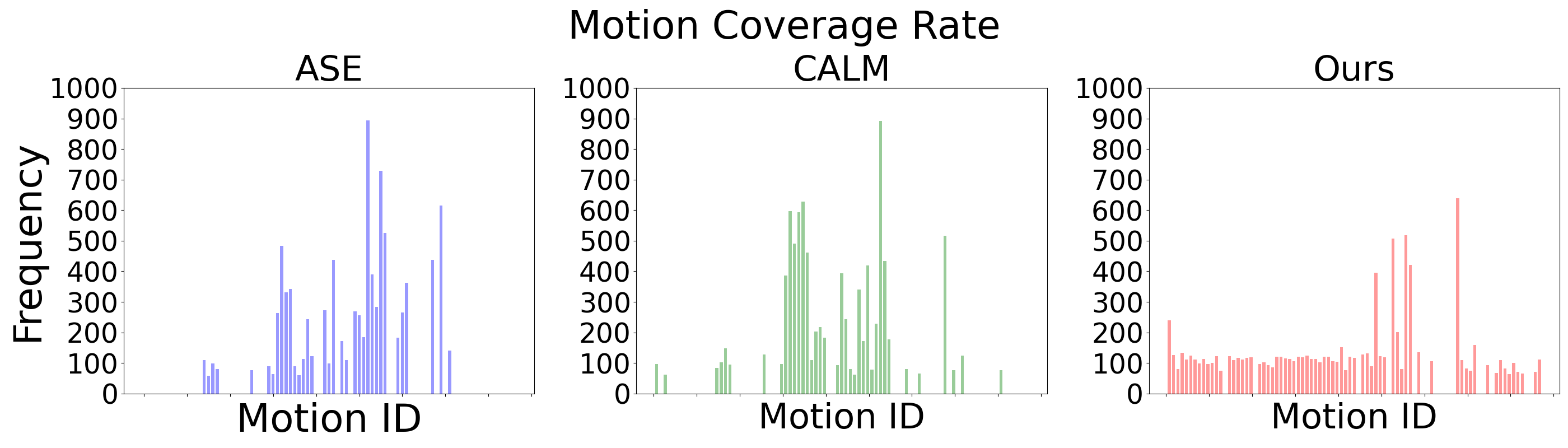}
  \put(33,-3){(b) Filtering rate $= 50\%$.}
\end{overpic}
  \makebox[\linewidth][c]{}\\
\caption{Frequencies at which the low-level policy produces motions that match all 87 individual clips. We show distributions produced by the filtering rate of 0\% and 50\% here. Compared to ASE and CALM, our method produces diverse motions that much more evenly cover all reference clips.}
\label{fig:motion_covrage_bar}
\end{figure}

Figure~\ref{fig:motion_covrage_cutoff_curve} presents the motion coverage rate under different \filteringrates. We can see although ASE achieves a competitive coverage rate to ours when no filtering is applied, its coverage dramatically drops to $66\%$, $57\%$, and $40\%$ at the filtering rate of $10\%$, $20\%$, and $50\%$, respectively. The coverage rate of CALM drops from $84\%$ to $71\%$, and $43\%$ at the filtering rate of $10\%$, $20\%$, and $50\%$. These results indicate a serious unbalance of the motion coverage, i.e., many motion clips are matched only a few times, possibly due to stochastic factors existing in the randomly generated trajectories. In contrast, our model produces consistently high motion coverage rates at different \filteringrates, indicating that all motion clips are matched rather evenly by the randomly generated trajectories. This is further evidenced by 
Figure~\ref{fig:motion_covrage_bar}, which records the frequencies at which $\policy$ produces trajectories that match each motion clip in the dataset across 10,000 trajectories. 
We also evaluate on the larger Composite Skills dataset. For a fair comparison, we conducted the evaluation on the Composite Skills dataset while maintaining consistent SRF settings for baseline comparisons. The results are as follows: a) With SRF, our approach achieved a filtered coverage rate of $82\%$, outperforming CALM with $58\%$ and ASE with $51\%$. b) Without SRF, our approach achieved a filtered coverage rate of $80\%$, surpassing CALM with $55\%$ and ASE with $44\%$. We observed that the inclusion of SRF positively impacted the performance of all models in terms of motion coverage. Notably, the increase in motion coverage was particularly evident for highly dynamic, agile, and stylized motions such as ballet, sidekick, and zombie walks. Despite these improvements, our model continues to excel in learning extensive and complex skills compared with CALM and ASE, which underscores the effectiveness of learning conditional skill embeddings and other key design elements in our framework. See more details in Appendix~\ref{ap:scalability}. 

\paragraph{Fréchet Inception Distance}
Following~\cite{hassan2021stochastic, wang2022towards}, we further measure the similarity between the distribution of generated motions and that of reference motions using Fréchet Inception Distance. The distance is computed using the character state of each frame. We report FID scores computed at three different levels: per frame, per transition (2 frames), and per clip (30 frames). As shown in Table~\ref{tab:FID}, our model achieves lower FID, indicating motions produced from \name are closer to the distribution of reference motions.
\begin{table}[t!]
  \caption{Fréchet Inception Distance (lower is better) comparison. A lower FID indicates that generated motions are closer to the reference distribution.}
  \label{tab:FID}
  \footnotesize
\begin{tabular}{lcccc}
    \toprule
 Input & \#Frame & ASE & CALM  & Ours   \\
    \midrule
    Per-frame & $1$& $28.8$  & $30.1$ & $\textbf{16.5}$ \\
    Per-transition & $2$& $72.3$  & $69.1$ & $\textbf{47.4}$\\
    Per-clip & $30$& $1969.8$ & $1874.6$  & $\textbf{1742.5}$\\
  \bottomrule
\end{tabular}
\end{table}

\begin{figure*}[]
\centering 
\includegraphics[width=\linewidth]{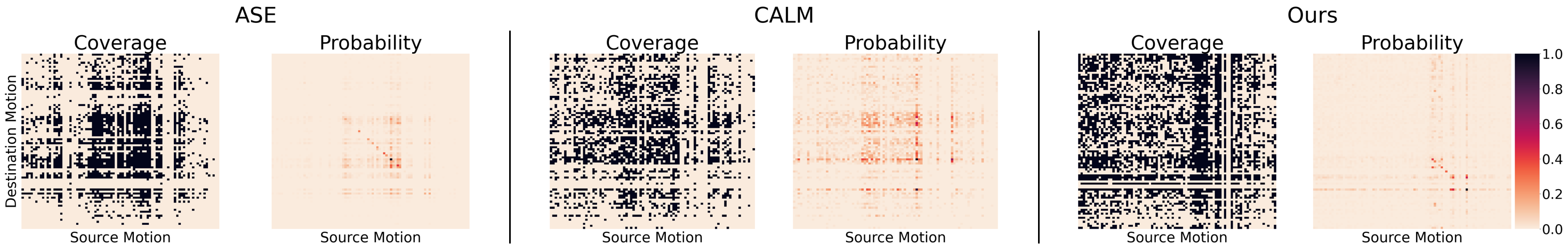}
\caption{Transition coverage and probability between different skills. We show all 87 motion clips in the Sword\&Shield dataset. Our model achieves a higher coverage rate and more even transition probability distribution compared with ASE and CALM.}
\label{fig:transition_coverage_rate}
\vspace{1mm}
\end{figure*}

\paragraph{Skill Transition Coverage}
It is important that the low-level policy can learn to transition between various skills to perform composed and sequenced skills in complex tasks. To evaluate the model’s capability to transition between different skills, we generate transition trajectories by conditioning on two pairs of random condition signals $p_1 = (\skilllabel, \skilllatent)$ and $ p_2 = (\skilllabel', \skilllatent')$ per trajectory. A transition trajectory is generated by first conditioning on $p_1$ for $200$ time steps, then is conditioned on $p_2$ for another $200$ time steps. Then, these two sub-trajectories are used to separately match in the dataset, using Equation~\ref{eq:filter_coverage_rate}, to identify a source motion  (denoted as source motion $\motionclip_S$) and a destination motion  (denoted as destination motion $\motionclip_D$). We repeat this process for $10,000$ transition trajectories and record the transition coverage, as well as the probability between each pair of motion clips.
We compare our model with ASE and CALM, where a transition trajectory is generated using two random skill codes. The transition coverage and probability result is shown in Figure~\ref{fig:transition_coverage_rate}, where \name produces a denser connection of each possible transition and the transition coverage is distributed more balanced compared with ASE and CALM. Furthermore, we report the \textit{transition coverage rate}
$ 
= \frac{\text{\#transitions from model}}{\text{\#all possible transitions}}$,
on which our model~(44.3\%) outperforms ASE~(25.4\%) and CALM~(28.4\%) by a margin of 74.4\% and 55.9\%, respectively.

\paragraph{Motion Diversity}
We evaluate the diversity of motions produced by the low-level model.  
Following~\cite{wang2022towards, lu2022action, hassan2021stochastic}
We adopt the Average Pairwise Distance (APD) to measure the diversity of a set of generated motion sequences.
Specifically, given a set of generated motion sequences $\refmotiondataset = \{ \motionclip_i\}$ where each motion clip $\motionclip_i$ contains $\framenum$ frames, the APD is computed as:
\begin{equation}
    APD(\refmotiondataset) = \frac{1}{N(N-1)} \sum_{i=1}^N  \sum_{j\ne i}^N(\sum_{t=1}^\framenum(\Vert \states^{i}_t - \states^{j}_t\Vert^2))^{\frac{1}{2}},
\end{equation}
where $s_t^i \in M_i$ is a state in a motion clip $M_i$ and $N$ is the number of generated sequences. A larger APD indicates a more diverse set of motion sequences. We compare with ASE and CALM on the mean and standard deviation of this metric 10 times to investigate the diversity across all generated motions. For each time, we test with $N = 10,000$ sequences that are generated by ASE and CALM conditioned on random latent codes and by our model conditioned on random skill labels and latent codes. As a result, our model produces a higher APD score ($160.4 \pm 1.77$), which indicates higher diversity of generated motions compared to ASE ($145.4 \pm 2.11$) as well as CALM ($152.7 \pm 1.86$).

\begin{figure*}[]
\centering
\includegraphics[width=\linewidth]{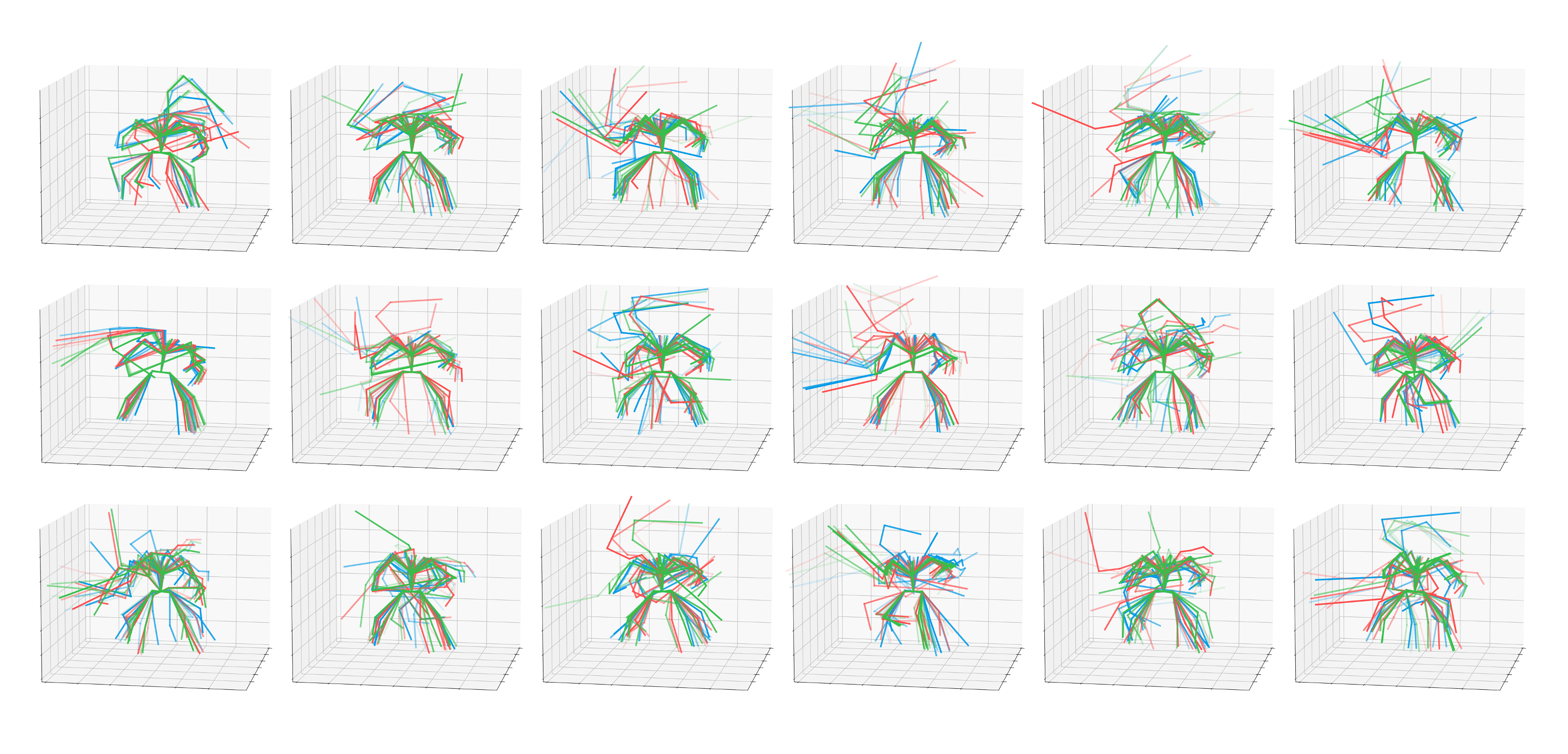}
\vspace{-10mm}
\caption{Motion variations (three colored in red, blue, and green are shown) are shown for each skill category. We test with the Sword\&Shield dataset.}
\label{fig:motion_diversity_local}
\vspace{4mm}
\end{figure*}

Moreover, we conduct qualitative evaluations of the motion diversity of our model: 
(i) \emph{Global root trajectory}.
We visualize the behaviors produced by random motions. Figure~\ref{fig:motion_diversity_root_motion} illustrates the root trajectories produced by different skill labels and latent codes. All motions are generated from the same initial idle state.  We generated $100$ trajectories, with each containing $300$ time steps,  for each top-$10$ skill (ranked by APD of the motion within the skill category). 
(ii)\emph{Local motion diversity}. 
We investigate the diversity of motions under each skill category by fixing the skill label $\skilllabel$ and randomizing the latent code $\skilllatent$. As demonstrated in Figure~\ref{fig:motion_diversity_local}, the motions produced with each skill label exhibit local variations while still conforming to the general characteristics of each skill. More qualitative results are presented in the supplementary video.

\begin{figure}[!t]
\centering
\begin{overpic}[width=0.9\linewidth]{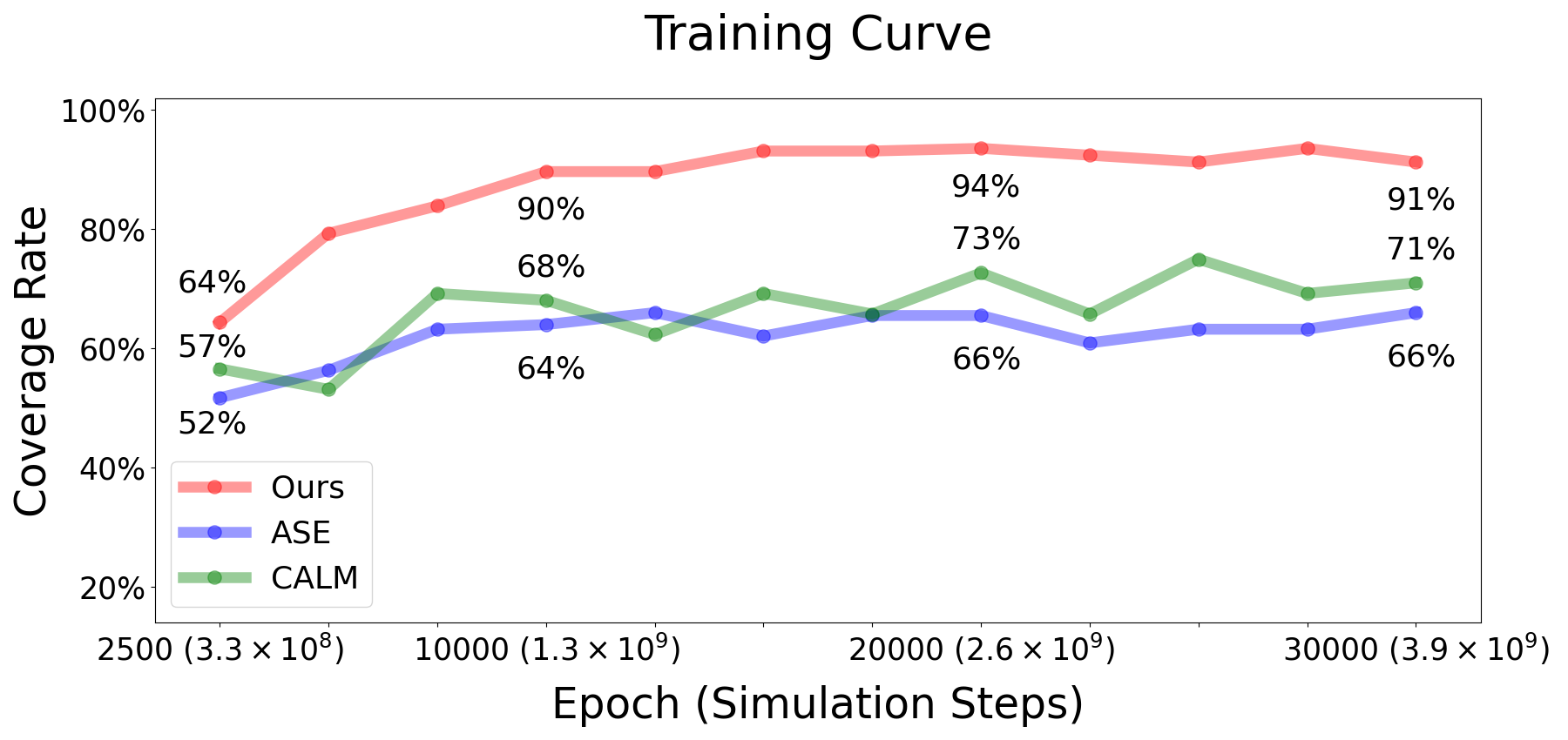}
\end{overpic}
\caption{
Comparison of training effectiveness and efficiency. We plot the filtered coverage rate (filtering rate = 10\%) w.r.t the training course.}
\vspace{-3mm}
\label{fig:training_convergence}
\end{figure}

\paragraph{Learning Efficiency and Effectiveness}
We show that learning a structured latent space via conditional adversarial imitation learning not only offers an explicit skill control handle but also greatly improves the effectiveness and efficiency of the training course. In Figure~\ref{fig:training_convergence}, our model can effectively cover around 91\% of the reference motions in the dataset within just 30,000 epochs, whereas CALM and ASE converge to only a coverage rate of 71\% and 66\% and 
barely improves with more epochs. We conjecture that the instability of CALM's coverage rate may be caused by the dynamically changing latent codes produced by the encoder during training.

\subsection{Ablation Study}
\label{sec:ablation_study}
\begin{figure*}[!tbp]
  \centering
  \begin{minipage}[b]{0.49\textwidth}
      \includegraphics[width=\textwidth]{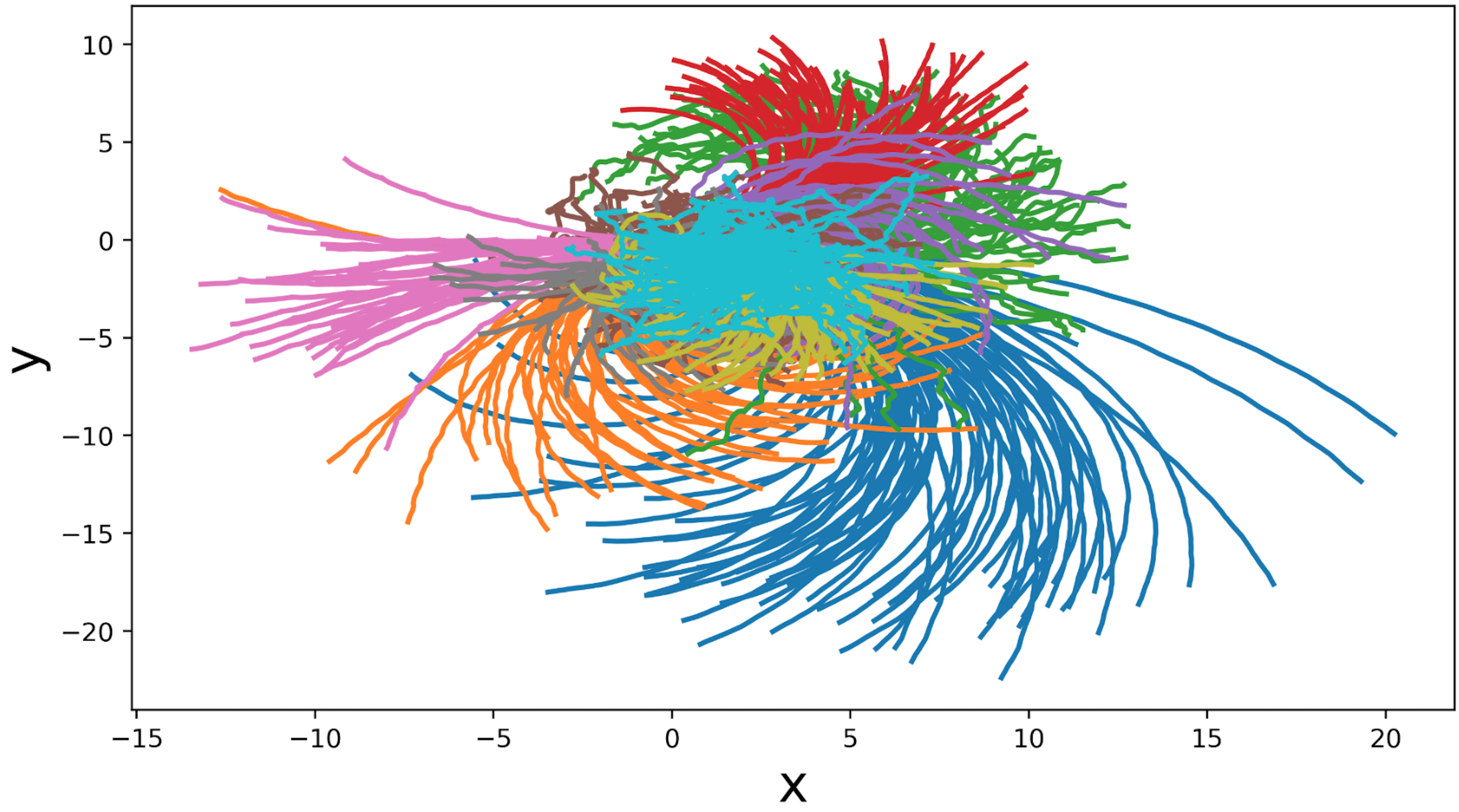}
    \vspace{-8mm}
    \caption{Visualization of trajectories of the character’s root produced by random exploration with the low-level model. The trajectories are generated from random skill labels and latent codes. Starting from the same idle states, the character is able to move in different directions with realistic motions.}
    \label{fig:motion_diversity_root_motion}
  \end{minipage}
  \hfill
  \begin{minipage}[b]{0.495\textwidth}
    \includegraphics[width=\textwidth]{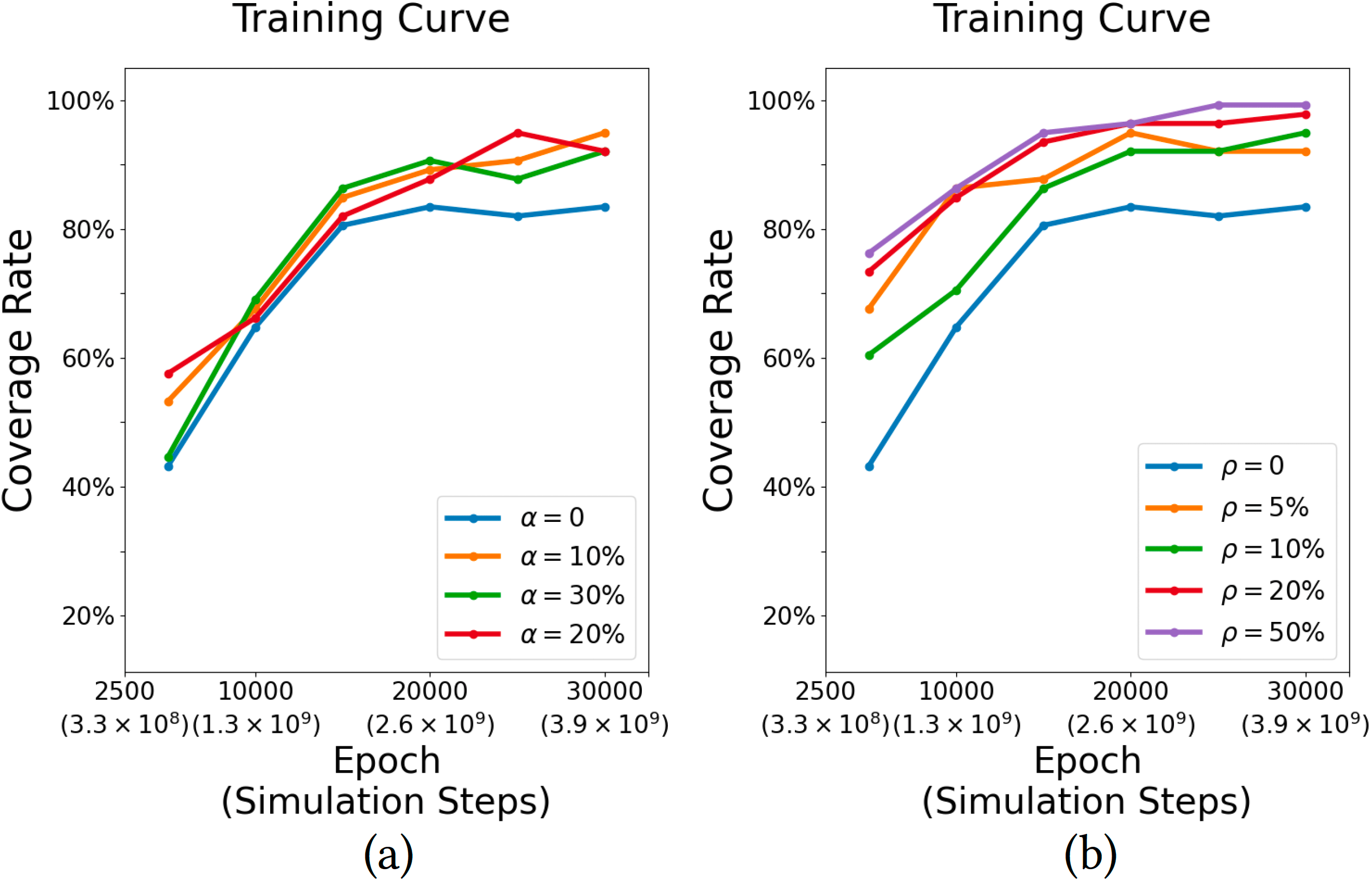}
    \caption{Influence of (a) the Focal Skill Sampling and (b) Element-wise Feature Masking to the training \ZY{of the low-level model}.}
    \label{fig:ablation_convergence}
  \end{minipage}
\end{figure*}

\paragraph{Focal Skill Sampling}
We compare with the baseline model that is trained without the focal skill sampling and those trained with different update rates $\forgetrate$. Figure~\ref{fig:ablation_convergence}~(a) shows the motion coverage rate that these models achieve during the training course. In general, models with focal skill sampling outperform the one where the module is ablated, converging to higher coverage rates. In our experiments, the update rate is set to $\forgetrate=20\%$ by default.

\paragraph{Skeletal Residual Force} 
We investigate the impact of Skeletal Residual Forces~(SRF) on the embedding of skills associated with agile movements. A qualitative comparison is presented in the supplementary video, demonstrating that SRF facilitates learning more agile motions. 
\ZY{We noticed that, while SRF is important in learning agile movements, it can compromise physical accuracy. So we introduced regularization ($r_{f} = \exp(-\sum^{J-2}j\bm{\xi}{j=1})$) to residual forces, ensuring their utilization only when necessary, for which we investigate in the following. We found the average residual forces (L2-norm of the force magnitude) across $17$ joints in 200-time steps of $1024$ trajectories generated by randomizing the skill label $\skilllabel$ and the latent code $\skilllatent$ during test time amount to only $0.842\%$ of the internal force derived from PD target control. 
This represents a small percentage and causes only minor deviations from physical correctness.
}

\paragraph{Element-wise Feature Masking}
Empirically, we found that this simple yet effective element-wise feature masking not only improved the motion diversity produced under each skill category with APD increased from $150.4 \pm 1.37$ to $160.2 \pm 1.23$ (measured by $10$ times) but also improved training efficiency, which can be observed in Figure~\ref{fig:ablation_convergence}~(b). We found setting the random probability $\maskingrate$ to large values leads to jittering; See the supplementary video. Thus, by default, we use $\maskingrate = 20\%$, which is a good trade-off in practice.

\subsection{Interactive Character Animation}
\label{sec:res_high_level_control}
\begin{figure*}
\centering
\includegraphics[width=\linewidth]{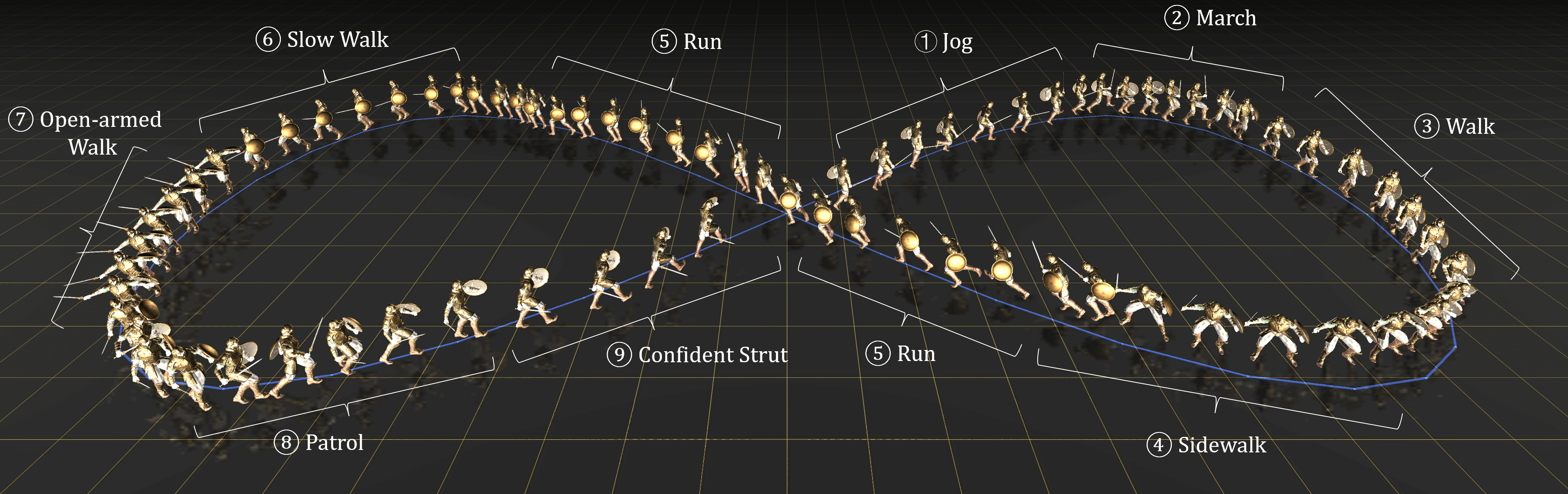}
\caption{Path-follower: the character faithfully follows user-specified paths (blue path) under various specified skills, including jog, march, walk, sidewalk, run, slow walk, open-armed walk, patrol, and confident strut; More results are presented in the supplementary video.}
\label{fig:application_path_following}
\vspace{4mm}
\end{figure*}

\begin{figure*}
\centering
\includegraphics[width=\linewidth]{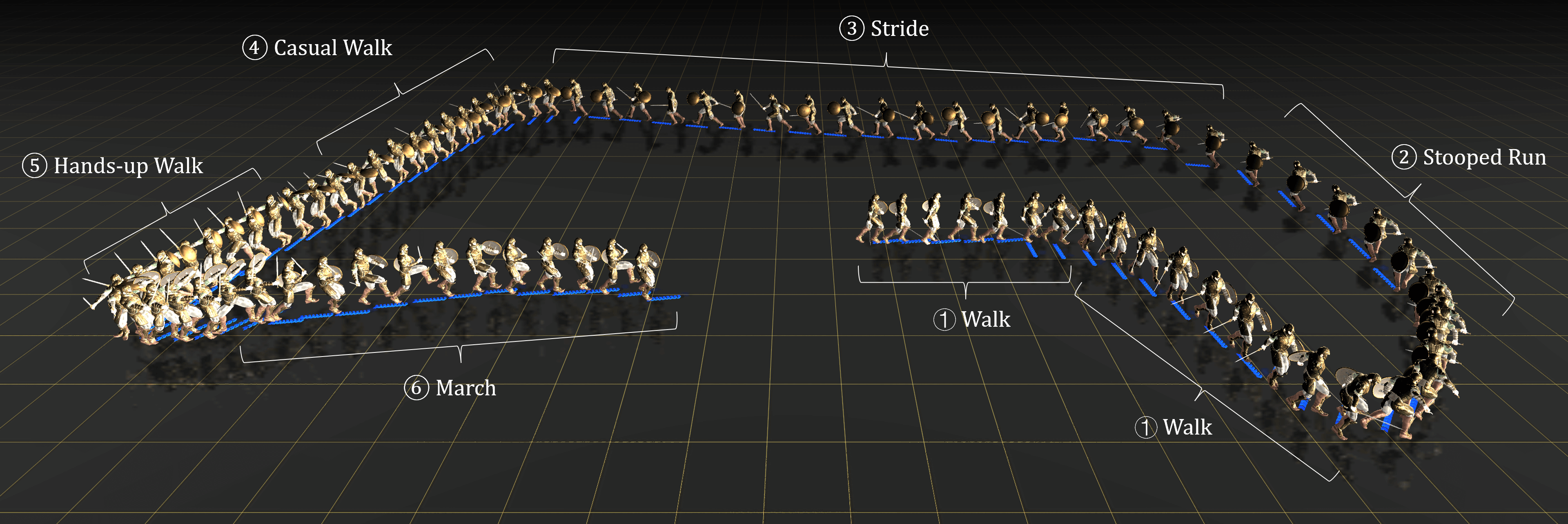}
\caption{Directional Control: the character faithfully follows user-specified local moving direction (blue arrow) under various specified skills, including walk, stooped run, stride, casual walk, hands-up walk, and march; See more results in the supplementary video.}
\label{fig:application_directional_control}
\vspace{4mm}
\end{figure*}

\begin{figure*}
\centering
\includegraphics[width=\linewidth]{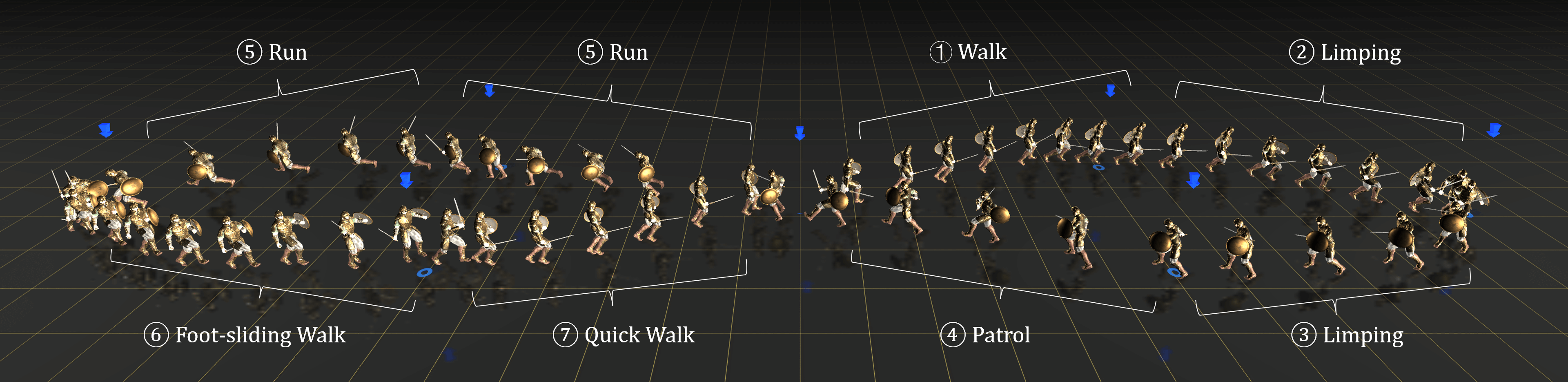}\caption{Character Re-locating: the character re-locates to a user-specified target location (visualized by the blue disks on the ground and arrows in the air) with different specified skills, including walk, limping, patrol, run, foot-sliding walk, and quick walk; More results can be found in the supplementary video.}
\label{fig:application_location}
\end{figure*}

We evaluate the efficacy of our framework in interactive character animation. Since the Sword\&Shield dataset contains only two locomotion skills, i.e., walking and running, we pre-train the low-level conditional model and those interactive controller policies on the larger Composite Skills dataset containing richer skills. We present interactive character animations in various ways realized with the learned conditional model and interactive controller policies, 
including \emph{path-follower}, \emph{directional control} and \emph{character re-locating} with the desired skill explicitly specified by the user. Figure~\ref{fig:application_path_following} depicts characters faithfully following user-specified paths under various specified skills. 
Figure~\ref{fig:application_directional_control} demonstrates dynamic user control over the moving direction with desired skills akin to those in video games, while Figure~\ref{fig:application_location} displays relocating the character to a specified location with desired skills. Additional results can be found in the supplementary video. We believe these features are valuable for video game and animation production.

\section{Discussion and Conclusion}
In this work, we introduce \name, an efficient and effective framework for learning Conditional Adversarial Skill Embedding for physics-based characters. The key idea is dividing the repertoire into homogeneous sub-sets and conquering them for learning conditional behavior distribution. Consequently, \name outperforms state-of-the-art methods, enabling characters to master diverse motor skills efficiently. Notably, skill-conditioned imitation learning naturally offers explicit control over the embedded skills. We demonstrate the application of such explicit control handles in controllable character animation in various ways, showing its superior practical value.

Despite its remarkable advantages, we note a few shortcomings. We are aware of some artifacts remaining in reproducing some CMU skills, as shown in the supplementary video. This is due to two main reasons: First, GAN-based models often suffer from mode-collapse issues, although we have shown that conditional distribution learning could significantly alleviate this problem. Exploring other generative models like the diffusion model~\cite{song2020denoising, mdm22, shi2023controllable} and VQ-GAN~\cite{vqgan} may be beneficial in the future. Second, practical operations performed during experiments can contribute to the artifacts. For example, our simplified skeleton with $17$ joints, compared to the original CMU skeleton with $31$ joints, may limit the expression of agile motions, resulting in stiffness. Last, our framework is highly sample-intensive. The artifacts may imply unsaturated sampling during the PPO training and could be mitigated by more sufficient training. This can be supported by the increasing motion coverage rate over longer training time. 

While the incorporation of SRF enhances motion quality and is compatible with various simulation platforms~\cite{isaacgym, todorov2012mujoco, Coumans2015BulletPS}, it may not be applied to real-world setups. \ZY{We believe SRF is not the optimal solution for learning highly agile and complex motions in the simulation environment, which demands more research effort into innovative solutions that guarantee physical correctness.} Another limitation of \name is the reliance on the skill label of the motion clips. Although we have shown that an action recognition network could help with motion segmentation to a large extent (see Appendix~\ref{ap:skill_acquisition}), developing a fully automatic framework for embedding extensive motions from unstructured data remains challenging. 

Last, a more powerful low-level model could, in turn, pose challenges to the learning of high-level strategies for more complex tasks since the effective action space has become larger. Thus, it would be worth exploring training high-level policies that can leverage diverse and extensive skills embeddings more effectively and efficiently for empowering the simulated character with the intelligence to undertake more complex tasks in more complicated environments. For example, training warriors that can make full use of all diverse and extensive skills to win a contest.

\begin{acks}
The authors would like to thank Zeshi Yang, Jintao Lu, Zhenhua Song, Heyuan Yao and Mingyi Shi for the fruitful discussion and the anonymous reviewers for their valuable comments and suggestions. Taku Komura is supported by Meta Reality Labs, Innovation and Technology Commission~(Ref: ITS/319/21FP) and Research Grant Council of Hong Kong~(Ref: 17210222). Wenping Wang is supported by Research Grant Council of Hong Kong~(Ref: T45-205/21-N).
\end{acks}
\bibliographystyle{ACM-Reference-Format}
\bibliography{bibliography.bib}


\begin{thebibliography}{51}


\ifx \showCODEN    \undefined \def \showCODEN     #1{\unskip}     \fi
\ifx \showDOI      \undefined \def \showDOI       #1{#1}\fi
\ifx \showISBNx    \undefined \def \showISBNx     #1{\unskip}     \fi
\ifx \showISBNxiii \undefined \def \showISBNxiii  #1{\unskip}     \fi
\ifx \showISSN     \undefined \def \showISSN      #1{\unskip}     \fi
\ifx \showLCCN     \undefined \def \showLCCN      #1{\unskip}     \fi
\ifx \shownote     \undefined \def \shownote      #1{#1}          \fi
\ifx \showarticletitle \undefined \def \showarticletitle #1{#1}   \fi
\ifx \showURL      \undefined \def \showURL       {\relax}        \fi
\providecommand\bibfield[2]{#2}
\providecommand\bibinfo[2]{#2}
\providecommand\natexlab[1]{#1}
\providecommand\showeprint[2][]{arXiv:#2}

\bibitem[Bergamin et~al\mbox{.}(2019)]%
        {bergamin2019drecon}
\bibfield{author}{\bibinfo{person}{Kevin Bergamin}, \bibinfo{person}{Simon Clavet}, \bibinfo{person}{Daniel Holden}, {and} \bibinfo{person}{James~Richard Forbes}.} \bibinfo{year}{2019}\natexlab{}.
\newblock \showarticletitle{DReCon: data-driven responsive control of physics-based characters}.
\newblock \bibinfo{journal}{\emph{ACM Transactions On Graphics (TOG)}} \bibinfo{volume}{38}, \bibinfo{number}{6} (\bibinfo{year}{2019}), \bibinfo{pages}{1--11}.
\newblock


\bibitem[Cai et~al\mbox{.}(2022)]%
        {cai2022humman}
\bibfield{author}{\bibinfo{person}{Zhongang Cai}, \bibinfo{person}{Daxuan Ren}, \bibinfo{person}{Ailing Zeng}, \bibinfo{person}{Zhengyu Lin}, \bibinfo{person}{Tao Yu}, \bibinfo{person}{Wenjia Wang}, \bibinfo{person}{Xiangyu Fan}, \bibinfo{person}{Yang Gao}, \bibinfo{person}{Yifan Yu}, \bibinfo{person}{Liang Pan}, {et~al\mbox{.}}} \bibinfo{year}{2022}\natexlab{}.
\newblock \showarticletitle{Humman: Multi-modal 4d human dataset for versatile sensing and modeling}. In \bibinfo{booktitle}{\emph{European Conference on Computer Vision}}. Springer, \bibinfo{pages}{557--577}.
\newblock


\bibitem[CMU(2002)]%
        {CMU}
\bibfield{author}{\bibinfo{person}{CMU}.} \bibinfo{year}{2002}\natexlab{}.
\newblock \bibinfo{title}{{ CMU Graphics Lab Motion Capture Database}}.
\newblock \bibinfo{howpublished}{\url{http://mocap.cs.cmu.edu/}}.
\newblock


\bibitem[Coumans(2015)]%
        {Coumans2015BulletPS}
\bibfield{author}{\bibinfo{person}{Erwin Coumans}.} \bibinfo{year}{2015}\natexlab{}.
\newblock \showarticletitle{Bullet physics simulation}.
\newblock \bibinfo{journal}{\emph{ACM SIGGRAPH 2015 Courses}} (\bibinfo{year}{2015}).
\newblock


\bibitem[Dou et~al\mbox{.}(2022)]%
        {dou2022tore}
\bibfield{author}{\bibinfo{person}{Zhiyang Dou}, \bibinfo{person}{Qingxuan Wu}, \bibinfo{person}{Cheng Lin}, \bibinfo{person}{Zeyu Cao}, \bibinfo{person}{Qiangqiang Wu}, \bibinfo{person}{Weilin Wan}, \bibinfo{person}{Taku Komura}, {and} \bibinfo{person}{Wenping Wang}.} \bibinfo{year}{2022}\natexlab{}.
\newblock \showarticletitle{TORE: Token Reduction for Efficient Human Mesh Recovery with Transformer}.
\newblock \bibinfo{journal}{\emph{arXiv preprint arXiv:2211.10705}} (\bibinfo{year}{2022}).
\newblock


\bibitem[Duan et~al\mbox{.}(2022)]%
        {duan2022pyskl}
\bibfield{author}{\bibinfo{person}{Haodong Duan}, \bibinfo{person}{Jiaqi Wang}, \bibinfo{person}{Kai Chen}, {and} \bibinfo{person}{Dahua Lin}.} \bibinfo{year}{2022}\natexlab{}.
\newblock \showarticletitle{Pyskl: Towards good practices for skeleton action recognition}. In \bibinfo{booktitle}{\emph{Proceedings of the 30th ACM International Conference on Multimedia}}. \bibinfo{pages}{7351--7354}.
\newblock


\bibitem[Esser et~al\mbox{.}(2021)]%
        {vqgan}
\bibfield{author}{\bibinfo{person}{Patrick Esser}, \bibinfo{person}{Robin Rombach}, {and} \bibinfo{person}{Bjorn Ommer}.} \bibinfo{year}{2021}\natexlab{}.
\newblock \showarticletitle{Taming transformers for high-resolution image synthesis}. In \bibinfo{booktitle}{\emph{Proceedings of the IEEE/CVF conference on computer vision and pattern recognition}}. \bibinfo{pages}{12873--12883}.
\newblock


\bibitem[Fan et~al\mbox{.}(2019)]%
        {fan2019hybrid}
\bibfield{author}{\bibinfo{person}{Zhou Fan}, \bibinfo{person}{Rui Su}, \bibinfo{person}{Weinan Zhang}, {and} \bibinfo{person}{Yong Yu}.} \bibinfo{year}{2019}\natexlab{}.
\newblock \showarticletitle{Hybrid actor-critic reinforcement learning in parameterized action space}. In \bibinfo{booktitle}{\emph{Proceedings of the 28th International Joint Conference on Artificial Intelligence}}. \bibinfo{pages}{2279--2285}.
\newblock


\bibitem[Fussell et~al\mbox{.}(2021)]%
        {fussell2021supertrack}
\bibfield{author}{\bibinfo{person}{Levi Fussell}, \bibinfo{person}{Kevin Bergamin}, {and} \bibinfo{person}{Daniel Holden}.} \bibinfo{year}{2021}\natexlab{}.
\newblock \showarticletitle{Supertrack: Motion tracking for physically simulated characters using supervised learning}.
\newblock \bibinfo{journal}{\emph{ACM Transactions on Graphics (TOG)}} \bibinfo{volume}{40}, \bibinfo{number}{6} (\bibinfo{year}{2021}), \bibinfo{pages}{1--13}.
\newblock


\bibitem[Harvey et~al\mbox{.}(2020)]%
        {harvey2020lafan}
\bibfield{author}{\bibinfo{person}{F{\'e}lix~G Harvey}, \bibinfo{person}{Mike Yurick}, \bibinfo{person}{Derek Nowrouzezahrai}, {and} \bibinfo{person}{Christopher Pal}.} \bibinfo{year}{2020}\natexlab{}.
\newblock \showarticletitle{Robust motion in-betweening}.
\newblock \bibinfo{journal}{\emph{ACM Transactions on Graphics (TOG)}} \bibinfo{volume}{39}, \bibinfo{number}{4} (\bibinfo{year}{2020}), \bibinfo{pages}{60--1}.
\newblock


\bibitem[Hassan et~al\mbox{.}(2021)]%
        {hassan2021stochastic}
\bibfield{author}{\bibinfo{person}{Mohamed Hassan}, \bibinfo{person}{Duygu Ceylan}, \bibinfo{person}{Ruben Villegas}, \bibinfo{person}{Jun Saito}, \bibinfo{person}{Jimei Yang}, \bibinfo{person}{Yi Zhou}, {and} \bibinfo{person}{Michael~J Black}.} \bibinfo{year}{2021}\natexlab{}.
\newblock \showarticletitle{Stochastic scene-aware motion prediction}. In \bibinfo{booktitle}{\emph{Proceedings of the IEEE/CVF International Conference on Computer Vision}}. \bibinfo{pages}{11374--11384}.
\newblock


\bibitem[Ho and Ermon(2016)]%
        {ho2016generative}
\bibfield{author}{\bibinfo{person}{Jonathan Ho} {and} \bibinfo{person}{Stefano Ermon}.} \bibinfo{year}{2016}\natexlab{}.
\newblock \showarticletitle{Generative adversarial imitation learning}.
\newblock \bibinfo{journal}{\emph{Advances in neural information processing systems}}  \bibinfo{volume}{29} (\bibinfo{year}{2016}).
\newblock


\bibitem[Jang et~al\mbox{.}(2016)]%
        {jang2016categorical}
\bibfield{author}{\bibinfo{person}{Eric Jang}, \bibinfo{person}{Shixiang Gu}, {and} \bibinfo{person}{Ben Poole}.} \bibinfo{year}{2016}\natexlab{}.
\newblock \showarticletitle{Categorical reparameterization with gumbel-softmax}.
\newblock \bibinfo{journal}{\emph{arXiv preprint arXiv:1611.01144}} (\bibinfo{year}{2016}).
\newblock


\bibitem[Juravsky et~al\mbox{.}(2022)]%
        {juravsky2022padl}
\bibfield{author}{\bibinfo{person}{Jordan Juravsky}, \bibinfo{person}{Yunrong Guo}, \bibinfo{person}{Sanja Fidler}, {and} \bibinfo{person}{Xue~Bin Peng}.} \bibinfo{year}{2022}\natexlab{}.
\newblock \showarticletitle{PADL: Language-Directed Physics-Based Character Control}. In \bibinfo{booktitle}{\emph{SIGGRAPH Asia 2022 Conference Papers}}. \bibinfo{pages}{1--9}.
\newblock


\bibitem[Li et~al\mbox{.}(2021a)]%
        {li2021hyar}
\bibfield{author}{\bibinfo{person}{Boyan Li}, \bibinfo{person}{Hongyao Tang}, \bibinfo{person}{YAN ZHENG}, \bibinfo{person}{HAO Jianye}, \bibinfo{person}{Pengyi Li}, \bibinfo{person}{Zhen Wang}, \bibinfo{person}{Zhaopeng Meng}, {and} \bibinfo{person}{LI Wang}.} \bibinfo{year}{2021}\natexlab{a}.
\newblock \showarticletitle{HyAR: Addressing Discrete-Continuous Action Reinforcement Learning via Hybrid Action Representation}. In \bibinfo{booktitle}{\emph{International Conference on Learning Representations}}.
\newblock


\bibitem[Li et~al\mbox{.}(2023)]%
        {li2023niki}
\bibfield{author}{\bibinfo{person}{Jiefeng Li}, \bibinfo{person}{Siyuan Bian}, \bibinfo{person}{Qi Liu}, \bibinfo{person}{Jiasheng Tang}, \bibinfo{person}{Fan Wang}, {and} \bibinfo{person}{Cewu Lu}.} \bibinfo{year}{2023}\natexlab{}.
\newblock \showarticletitle{NIKI: Neural Inverse Kinematics with Invertible Neural Networks for 3D Human Pose and Shape Estimation}.
\newblock \bibinfo{journal}{\emph{arXiv preprint arXiv:2305.08590}} (\bibinfo{year}{2023}).
\newblock


\bibitem[Li et~al\mbox{.}(2021b)]%
        {li2021hybrik}
\bibfield{author}{\bibinfo{person}{Jiefeng Li}, \bibinfo{person}{Chao Xu}, \bibinfo{person}{Zhicun Chen}, \bibinfo{person}{Siyuan Bian}, \bibinfo{person}{Lixin Yang}, {and} \bibinfo{person}{Cewu Lu}.} \bibinfo{year}{2021}\natexlab{b}.
\newblock \showarticletitle{Hybrik: A hybrid analytical-neural inverse kinematics solution for 3d human pose and shape estimation}. In \bibinfo{booktitle}{\emph{Proceedings of the IEEE/CVF conference on computer vision and pattern recognition}}. \bibinfo{pages}{3383--3393}.
\newblock


\bibitem[Liu et~al\mbox{.}(2020)]%
        {liu2020ntu}
\bibfield{author}{\bibinfo{person}{Jun Liu}, \bibinfo{person}{Amir Shahroudy}, \bibinfo{person}{Mauricio Perez}, \bibinfo{person}{Gang Wang}, \bibinfo{person}{Ling-Yu Duan}, {and} \bibinfo{person}{Alex~C Kot}.} \bibinfo{year}{2020}\natexlab{}.
\newblock \showarticletitle{NTU RGB+D 120: A large-scale benchmark for 3D human activity understanding}.
\newblock \bibinfo{journal}{\emph{IEEE Transactions on Pattern Analysis and Machine Intelligence}} \bibinfo{volume}{42}, \bibinfo{number}{10} (\bibinfo{year}{2020}), \bibinfo{pages}{2684--2701}.
\newblock


\bibitem[Liu and Hodgins(2018)]%
        {liu2018learning}
\bibfield{author}{\bibinfo{person}{Libin Liu} {and} \bibinfo{person}{Jessica Hodgins}.} \bibinfo{year}{2018}\natexlab{}.
\newblock \showarticletitle{Learning basketball dribbling skills using trajectory optimization and deep reinforcement learning}.
\newblock \bibinfo{journal}{\emph{ACM Transactions on Graphics (TOG)}} \bibinfo{volume}{37}, \bibinfo{number}{4} (\bibinfo{year}{2018}), \bibinfo{pages}{1--14}.
\newblock


\bibitem[Loper et~al\mbox{.}(2015)]%
        {loper2015smpl}
\bibfield{author}{\bibinfo{person}{Matthew Loper}, \bibinfo{person}{Naureen Mahmood}, \bibinfo{person}{Javier Romero}, \bibinfo{person}{Gerard Pons-Moll}, {and} \bibinfo{person}{Michael~J Black}.} \bibinfo{year}{2015}\natexlab{}.
\newblock \showarticletitle{SMPL: A skinned multi-person linear model}.
\newblock \bibinfo{journal}{\emph{ACM transactions on graphics (TOG)}} \bibinfo{volume}{34}, \bibinfo{number}{6} (\bibinfo{year}{2015}), \bibinfo{pages}{1--16}.
\newblock


\bibitem[Lu et~al\mbox{.}(2022)]%
        {lu2022action}
\bibfield{author}{\bibinfo{person}{Qiujing Lu}, \bibinfo{person}{Yipeng Zhang}, \bibinfo{person}{Mingjian Lu}, {and} \bibinfo{person}{Vwani Roychowdhury}.} \bibinfo{year}{2022}\natexlab{}.
\newblock \showarticletitle{Action-conditioned On-demand Motion Generation}. In \bibinfo{booktitle}{\emph{Proceedings of the 30th ACM International Conference on Multimedia}}. \bibinfo{pages}{2249--2257}.
\newblock


\bibitem[Mahmood et~al\mbox{.}(2019)]%
        {mahmood2019amass}
\bibfield{author}{\bibinfo{person}{Naureen Mahmood}, \bibinfo{person}{Nima Ghorbani}, \bibinfo{person}{Nikolaus~F Troje}, \bibinfo{person}{Gerard Pons-Moll}, {and} \bibinfo{person}{Michael~J Black}.} \bibinfo{year}{2019}\natexlab{}.
\newblock \showarticletitle{AMASS: Archive of motion capture as surface shapes}. In \bibinfo{booktitle}{\emph{Proceedings of the IEEE/CVF international conference on computer vision}}. \bibinfo{pages}{5442--5451}.
\newblock


\bibitem[Makoviychuk et~al\mbox{.}(2021)]%
        {isaacgym}
\bibfield{author}{\bibinfo{person}{Viktor Makoviychuk}, \bibinfo{person}{Lukasz Wawrzyniak}, \bibinfo{person}{Yunrong Guo}, \bibinfo{person}{Michelle Lu}, \bibinfo{person}{Kier Storey}, \bibinfo{person}{Miles Macklin}, \bibinfo{person}{David Hoeller}, \bibinfo{person}{Nikita Rudin}, \bibinfo{person}{Arthur Allshire}, \bibinfo{person}{Ankur Handa}, {and} \bibinfo{person}{Gavriel State}.} \bibinfo{year}{2021}\natexlab{}.
\newblock \showarticletitle{Isaac Gym: High Performance {GPU} Based Physics Simulation For Robot Learning}. In \bibinfo{booktitle}{\emph{Proceedings of the Neural Information Processing Systems Track on Datasets and Benchmarks 1, NeurIPS Datasets and Benchmarks 2021, December 2021, virtual}}, \bibfield{editor}{\bibinfo{person}{Joaquin Vanschoren} {and} \bibinfo{person}{Sai{-}Kit Yeung}} (Eds.).
\newblock
\urldef\tempurl%
\url{https://datasets-benchmarks-proceedings.neurips.cc/paper/2021/hash/28dd2c7955ce926456240b2ff0100bde-Abstract-round2.html}
\showURL{%
\tempurl}


\bibitem[Merel et~al\mbox{.}(2018)]%
        {merel2018neural}
\bibfield{author}{\bibinfo{person}{Josh Merel}, \bibinfo{person}{Leonard Hasenclever}, \bibinfo{person}{Alexandre Galashov}, \bibinfo{person}{Arun Ahuja}, \bibinfo{person}{Vu Pham}, \bibinfo{person}{Greg Wayne}, \bibinfo{person}{Yee~Whye Teh}, {and} \bibinfo{person}{Nicolas Heess}.} \bibinfo{year}{2018}\natexlab{}.
\newblock \showarticletitle{Neural probabilistic motor primitives for humanoid control}.
\newblock \bibinfo{journal}{\emph{arXiv preprint arXiv:1811.11711}} (\bibinfo{year}{2018}).
\newblock


\bibitem[Merel et~al\mbox{.}(2020)]%
        {merel2020catchandcarry}
\bibfield{author}{\bibinfo{person}{Josh Merel}, \bibinfo{person}{Saran Tunyasuvunakool}, \bibinfo{person}{Arun Ahuja}, \bibinfo{person}{Yuval Tassa}, \bibinfo{person}{Leonard Hasenclever}, \bibinfo{person}{Vu Pham}, \bibinfo{person}{Tom Erez}, \bibinfo{person}{Greg Wayne}, {and} \bibinfo{person}{Nicolas Heess}.} \bibinfo{year}{2020}\natexlab{}.
\newblock \showarticletitle{Catch \& Carry: reusable neural controllers for vision-guided whole-body tasks}.
\newblock \bibinfo{journal}{\emph{ACM Transactions on Graphics (TOG)}} \bibinfo{volume}{39}, \bibinfo{number}{4} (\bibinfo{year}{2020}), \bibinfo{pages}{39--1}.
\newblock


\bibitem[Nair et~al\mbox{.}(2018)]%
        {nair2018overcoming}
\bibfield{author}{\bibinfo{person}{Ashvin Nair}, \bibinfo{person}{Bob McGrew}, \bibinfo{person}{Marcin Andrychowicz}, \bibinfo{person}{Wojciech Zaremba}, {and} \bibinfo{person}{Pieter Abbeel}.} \bibinfo{year}{2018}\natexlab{}.
\newblock \showarticletitle{Overcoming exploration in reinforcement learning with demonstrations}. In \bibinfo{booktitle}{\emph{2018 IEEE international conference on robotics and automation (ICRA)}}. IEEE, \bibinfo{pages}{6292--6299}.
\newblock


\bibitem[Park et~al\mbox{.}(2019)]%
        {park2019learning}
\bibfield{author}{\bibinfo{person}{Soohwan Park}, \bibinfo{person}{Hoseok Ryu}, \bibinfo{person}{Seyoung Lee}, \bibinfo{person}{Sunmin Lee}, {and} \bibinfo{person}{Jehee Lee}.} \bibinfo{year}{2019}\natexlab{}.
\newblock \showarticletitle{Learning predict-and-simulate policies from unorganized human motion data}.
\newblock \bibinfo{journal}{\emph{ACM Transactions on Graphics (TOG)}} \bibinfo{volume}{38}, \bibinfo{number}{6} (\bibinfo{year}{2019}), \bibinfo{pages}{1--11}.
\newblock


\bibitem[Peng et~al\mbox{.}(2018)]%
        {peng2018deepmimic}
\bibfield{author}{\bibinfo{person}{Xue~Bin Peng}, \bibinfo{person}{Pieter Abbeel}, \bibinfo{person}{Sergey Levine}, {and} \bibinfo{person}{Michiel Van~de Panne}.} \bibinfo{year}{2018}\natexlab{}.
\newblock \showarticletitle{Deepmimic: Example-guided deep reinforcement learning of physics-based character skills}.
\newblock \bibinfo{journal}{\emph{ACM Transactions On Graphics (TOG)}} \bibinfo{volume}{37}, \bibinfo{number}{4} (\bibinfo{year}{2018}), \bibinfo{pages}{1--14}.
\newblock


\bibitem[Peng et~al\mbox{.}(2019)]%
        {peng2019mcp}
\bibfield{author}{\bibinfo{person}{Xue~Bin Peng}, \bibinfo{person}{Michael Chang}, \bibinfo{person}{Grace Zhang}, \bibinfo{person}{Pieter Abbeel}, {and} \bibinfo{person}{Sergey Levine}.} \bibinfo{year}{2019}\natexlab{}.
\newblock \showarticletitle{Mcp: Learning composable hierarchical control with multiplicative compositional policies}.
\newblock \bibinfo{journal}{\emph{Advances in Neural Information Processing Systems}}  \bibinfo{volume}{32} (\bibinfo{year}{2019}).
\newblock


\bibitem[Peng et~al\mbox{.}(2022)]%
        {ase}
\bibfield{author}{\bibinfo{person}{Xue~Bin Peng}, \bibinfo{person}{Yunrong Guo}, \bibinfo{person}{Lina Halper}, \bibinfo{person}{Sergey Levine}, {and} \bibinfo{person}{Sanja Fidler}.} \bibinfo{year}{2022}\natexlab{}.
\newblock \showarticletitle{ASE: Large-Scale Reusable Adversarial Skill Embeddings for Physically Simulated Characters}.
\newblock \bibinfo{journal}{\emph{ACM Transactions on Graphics (TOG)}} \bibinfo{volume}{41}, \bibinfo{number}{4}, Article \bibinfo{articleno}{94} (\bibinfo{date}{jul} \bibinfo{year}{2022}), \bibinfo{numpages}{17}~pages.
\newblock
\showISSN{0730-0301}
\urldef\tempurl%
\url{https://doi.org/10.1145/3528223.3530110}
\showDOI{\tempurl}


\bibitem[Peng et~al\mbox{.}(2021)]%
        {peng2021amp}
\bibfield{author}{\bibinfo{person}{Xue~Bin Peng}, \bibinfo{person}{Ze Ma}, \bibinfo{person}{Pieter Abbeel}, \bibinfo{person}{Sergey Levine}, {and} \bibinfo{person}{Angjoo Kanazawa}.} \bibinfo{year}{2021}\natexlab{}.
\newblock \showarticletitle{Amp: Adversarial motion priors for stylized physics-based character control}.
\newblock \bibinfo{journal}{\emph{ACM Transactions on Graphics (TOG)}} \bibinfo{volume}{40}, \bibinfo{number}{4} (\bibinfo{year}{2021}), \bibinfo{pages}{1--20}.
\newblock


\bibitem[Radford et~al\mbox{.}(2021)]%
        {clip}
\bibfield{author}{\bibinfo{person}{Alec Radford}, \bibinfo{person}{Jong~Wook Kim}, \bibinfo{person}{Chris Hallacy}, \bibinfo{person}{Aditya Ramesh}, \bibinfo{person}{Gabriel Goh}, \bibinfo{person}{Sandhini Agarwal}, \bibinfo{person}{Girish Sastry}, \bibinfo{person}{Amanda Askell}, \bibinfo{person}{Pamela Mishkin}, \bibinfo{person}{Jack Clark}, {et~al\mbox{.}}} \bibinfo{year}{2021}\natexlab{}.
\newblock \showarticletitle{Learning transferable visual models from natural language supervision}. In \bibinfo{booktitle}{\emph{International conference on machine learning}}. PMLR, \bibinfo{pages}{8748--8763}.
\newblock


\bibitem[Rajeswaran et~al\mbox{.}(2017)]%
        {rajeswaran2017learning}
\bibfield{author}{\bibinfo{person}{Aravind Rajeswaran}, \bibinfo{person}{Vikash Kumar}, \bibinfo{person}{Abhishek Gupta}, \bibinfo{person}{Giulia Vezzani}, \bibinfo{person}{John Schulman}, \bibinfo{person}{Emanuel Todorov}, {and} \bibinfo{person}{Sergey Levine}.} \bibinfo{year}{2017}\natexlab{}.
\newblock \showarticletitle{Learning complex dexterous manipulation with deep reinforcement learning and demonstrations}.
\newblock \bibinfo{journal}{\emph{arXiv preprint arXiv:1709.10087}} (\bibinfo{year}{2017}).
\newblock


\bibitem[Schulman et~al\mbox{.}(2017)]%
        {ppo}
\bibfield{author}{\bibinfo{person}{John Schulman}, \bibinfo{person}{Filip Wolski}, \bibinfo{person}{Prafulla Dhariwal}, \bibinfo{person}{Alec Radford}, {and} \bibinfo{person}{Oleg Klimov}.} \bibinfo{year}{2017}\natexlab{}.
\newblock \showarticletitle{Proximal Policy Optimization Algorithms}.
\newblock \bibinfo{journal}{\emph{CoRR}}  \bibinfo{volume}{abs/1707.06347} (\bibinfo{year}{2017}).
\newblock
\showeprint[arXiv]{1707.06347}
\urldef\tempurl%
\url{http://arxiv.org/abs/1707.06347}
\showURL{%
\tempurl}


\bibitem[SFU(2011)]%
        {SFU}
\bibfield{author}{\bibinfo{person}{NUS SFU}.} \bibinfo{year}{2011}\natexlab{}.
\newblock \bibinfo{title}{{SFU Motion Capture Database}}.
\newblock \bibinfo{howpublished}{\url{https://mocap.cs.sfu.ca/}}.
\newblock


\bibitem[Sharon and van~de Panne(2005)]%
        {sharon2005synthesis}
\bibfield{author}{\bibinfo{person}{Dana Sharon} {and} \bibinfo{person}{Michiel van~de Panne}.} \bibinfo{year}{2005}\natexlab{}.
\newblock \showarticletitle{Synthesis of controllers for stylized planar bipedal walking}. In \bibinfo{booktitle}{\emph{Proceedings of the 2005 IEEE International Conference on Robotics and Automation}}. IEEE, \bibinfo{pages}{2387--2392}.
\newblock


\bibitem[Shi et~al\mbox{.}(2020)]%
        {shi2020motionet}
\bibfield{author}{\bibinfo{person}{Mingyi Shi}, \bibinfo{person}{Kfir Aberman}, \bibinfo{person}{Andreas Aristidou}, \bibinfo{person}{Taku Komura}, \bibinfo{person}{Dani Lischinski}, \bibinfo{person}{Daniel Cohen-Or}, {and} \bibinfo{person}{Baoquan Chen}.} \bibinfo{year}{2020}\natexlab{}.
\newblock \showarticletitle{MotioNet: 3D Human Motion Reconstruction from Monocular Video with Skeleton Consistency}.
\newblock \bibinfo{journal}{\emph{arXiv preprint arXiv:2006.12075}} (\bibinfo{year}{2020}).
\newblock


\bibitem[Shi et~al\mbox{.}(2023)]%
        {shi2023controllable}
\bibfield{author}{\bibinfo{person}{Yi Shi}, \bibinfo{person}{Jingbo Wang}, \bibinfo{person}{Xuekun Jiang}, {and} \bibinfo{person}{Bo Dai}.} \bibinfo{year}{2023}\natexlab{}.
\newblock \showarticletitle{Controllable Motion Diffusion Model}.
\newblock \bibinfo{journal}{\emph{arXiv preprint arXiv:2306.00416}} (\bibinfo{year}{2023}).
\newblock


\bibitem[Song et~al\mbox{.}(2020)]%
        {song2020denoising}
\bibfield{author}{\bibinfo{person}{Jiaming Song}, \bibinfo{person}{Chenlin Meng}, {and} \bibinfo{person}{Stefano Ermon}.} \bibinfo{year}{2020}\natexlab{}.
\newblock \showarticletitle{Denoising Diffusion Implicit Models}. In \bibinfo{booktitle}{\emph{International Conference on Learning Representations}}.
\newblock


\bibitem[Tessler et~al\mbox{.}(2023)]%
        {calm}
\bibfield{author}{\bibinfo{person}{Chen Tessler}, \bibinfo{person}{Yoni Kasten}, \bibinfo{person}{Yunrong Guo}, \bibinfo{person}{Shie Mannor}, \bibinfo{person}{Gal Chechik}, {and} \bibinfo{person}{Xue~Bin Peng}.} \bibinfo{year}{2023}\natexlab{}.
\newblock \showarticletitle{CALM: Conditional Adversarial Latent Models for Directable Virtual Characters}.
\newblock \bibinfo{journal}{\emph{arXiv preprint arXiv:2305.02195}} (\bibinfo{year}{2023}).
\newblock


\bibitem[Tevet et~al\mbox{.}(2022)]%
        {mdm22}
\bibfield{author}{\bibinfo{person}{Guy Tevet}, \bibinfo{person}{Sigal Raab}, \bibinfo{person}{Brian Gordon}, \bibinfo{person}{Yonatan Shafir}, \bibinfo{person}{Daniel Cohen-Or}, {and} \bibinfo{person}{Amit~H Bermano}.} \bibinfo{year}{2022}\natexlab{}.
\newblock \showarticletitle{Human motion diffusion model}.
\newblock \bibinfo{journal}{\emph{arXiv preprint arXiv:2209.14916}} (\bibinfo{year}{2022}).
\newblock


\bibitem[Todorov et~al\mbox{.}(2012)]%
        {todorov2012mujoco}
\bibfield{author}{\bibinfo{person}{Emanuel Todorov}, \bibinfo{person}{Tom Erez}, {and} \bibinfo{person}{Yuval Tassa}.} \bibinfo{year}{2012}\natexlab{}.
\newblock \showarticletitle{Mujoco: A physics engine for model-based control}. In \bibinfo{booktitle}{\emph{2012 IEEE/RSJ international conference on intelligent robots and systems}}. IEEE, \bibinfo{pages}{5026--5033}.
\newblock


\bibitem[Tsuchida et~al\mbox{.}(2019)]%
        {tsuchida2019aist}
\bibfield{author}{\bibinfo{person}{Shuhei Tsuchida}, \bibinfo{person}{Satoru Fukayama}, \bibinfo{person}{Masahiro Hamasaki}, {and} \bibinfo{person}{Masataka Goto}.} \bibinfo{year}{2019}\natexlab{}.
\newblock \showarticletitle{AIST Dance Video Database: Multi-Genre, Multi-Dancer, and Multi-Camera Database for Dance Information Processing.}. In \bibinfo{booktitle}{\emph{ISMIR}}, Vol.~\bibinfo{volume}{1}. \bibinfo{pages}{6}.
\newblock


\bibitem[Wan et~al\mbox{.}(2021)]%
        {wan2021encoder}
\bibfield{author}{\bibinfo{person}{Ziniu Wan}, \bibinfo{person}{Zhengjia Li}, \bibinfo{person}{Maoqing Tian}, \bibinfo{person}{Jianbo Liu}, \bibinfo{person}{Shuai Yi}, {and} \bibinfo{person}{Hongsheng Li}.} \bibinfo{year}{2021}\natexlab{}.
\newblock \showarticletitle{Encoder-decoder with multi-level attention for 3D human shape and pose estimation}. In \bibinfo{booktitle}{\emph{Proceedings of the IEEE/CVF International Conference on Computer Vision}}. \bibinfo{pages}{13033--13042}.
\newblock


\bibitem[Wang et~al\mbox{.}(2022)]%
        {wang2022towards}
\bibfield{author}{\bibinfo{person}{Jingbo Wang}, \bibinfo{person}{Yu Rong}, \bibinfo{person}{Jingyuan Liu}, \bibinfo{person}{Sijie Yan}, \bibinfo{person}{Dahua Lin}, {and} \bibinfo{person}{Bo Dai}.} \bibinfo{year}{2022}\natexlab{}.
\newblock \showarticletitle{Towards Diverse and Natural Scene-aware 3D Human Motion Synthesis}. In \bibinfo{booktitle}{\emph{Proceedings of the IEEE/CVF Conference on Computer Vision and Pattern Recognition}}. \bibinfo{pages}{20460--20469}.
\newblock


\bibitem[Wang et~al\mbox{.}(2020)]%
        {wang2020unicon}
\bibfield{author}{\bibinfo{person}{Tingwu Wang}, \bibinfo{person}{Yunrong Guo}, \bibinfo{person}{Maria Shugrina}, {and} \bibinfo{person}{Sanja Fidler}.} \bibinfo{year}{2020}\natexlab{}.
\newblock \showarticletitle{Unicon: Universal neural controller for physics-based character motion}.
\newblock \bibinfo{journal}{\emph{arXiv preprint arXiv:2011.15119}} (\bibinfo{year}{2020}).
\newblock


\bibitem[Wang et~al\mbox{.}(2023)]%
        {wang2023zolly}
\bibfield{author}{\bibinfo{person}{Wenjia Wang}, \bibinfo{person}{Yongtao Ge}, \bibinfo{person}{Haiyi Mei}, \bibinfo{person}{Zhongang Cai}, \bibinfo{person}{Qingping Sun}, \bibinfo{person}{Yanjun Wang}, \bibinfo{person}{Chunhua Shen}, \bibinfo{person}{Lei Yang}, {and} \bibinfo{person}{Taku Komura}.} \bibinfo{year}{2023}\natexlab{}.
\newblock \showarticletitle{Zolly: Zoom Focal Length Correctly for Perspective-Distorted Human Mesh Reconstruction}.
\newblock \bibinfo{journal}{\emph{arXiv preprint arXiv:2303.13796}} (\bibinfo{year}{2023}).
\newblock


\bibitem[Won et~al\mbox{.}(2020)]%
        {won2020scalable}
\bibfield{author}{\bibinfo{person}{Jungdam Won}, \bibinfo{person}{Deepak Gopinath}, {and} \bibinfo{person}{Jessica Hodgins}.} \bibinfo{year}{2020}\natexlab{}.
\newblock \showarticletitle{A scalable approach to control diverse behaviors for physically simulated characters}.
\newblock \bibinfo{journal}{\emph{ACM Transactions on Graphics (TOG)}} \bibinfo{volume}{39}, \bibinfo{number}{4} (\bibinfo{year}{2020}), \bibinfo{pages}{33--1}.
\newblock


\bibitem[Won et~al\mbox{.}(2022)]%
        {cvae_meta}
\bibfield{author}{\bibinfo{person}{Jungdam Won}, \bibinfo{person}{Deepak Gopinath}, {and} \bibinfo{person}{Jessica Hodgins}.} \bibinfo{year}{2022}\natexlab{}.
\newblock \showarticletitle{Physics-based character controllers using conditional VAEs}.
\newblock \bibinfo{journal}{\emph{ACM Transactions on Graphics (TOG)}} \bibinfo{volume}{41}, \bibinfo{number}{4} (\bibinfo{year}{2022}), \bibinfo{pages}{1--12}.
\newblock


\bibitem[Yao et~al\mbox{.}(2022)]%
        {controlvae}
\bibfield{author}{\bibinfo{person}{Heyuan Yao}, \bibinfo{person}{Zhenhua Song}, \bibinfo{person}{Baoquan Chen}, {and} \bibinfo{person}{Libin Liu}.} \bibinfo{year}{2022}\natexlab{}.
\newblock \showarticletitle{ControlVAE: Model-Based Learning of Generative Controllers for Physics-Based Characters}.
\newblock \bibinfo{journal}{\emph{ACM Transactions on Graphics (TOG)}} \bibinfo{volume}{41}, \bibinfo{number}{6} (\bibinfo{year}{2022}), \bibinfo{pages}{1--16}.
\newblock


\bibitem[Yuan and Kitani(2020)]%
        {yuan2020residual}
\bibfield{author}{\bibinfo{person}{Ye Yuan} {and} \bibinfo{person}{Kris Kitani}.} \bibinfo{year}{2020}\natexlab{}.
\newblock \showarticletitle{Residual force control for agile human behavior imitation and extended motion synthesis}.
\newblock \bibinfo{journal}{\emph{Advances in Neural Information Processing Systems}}  \bibinfo{volume}{33} (\bibinfo{year}{2020}), \bibinfo{pages}{21763--21774}.
\newblock


\end{thebibliography}
\newpage

\clearpage 
\appendix
\renewcommand\thefigure{\Alph{section}\arabic{figure}}    
\renewcommand\thetable{\Alph{section}\arabic{table}}
\section{Appendix}

\subsection{Scalability}
\label{ap:scalability}
\begin{figure}[h]
\centering
\begin{overpic}[width=0.93\linewidth]{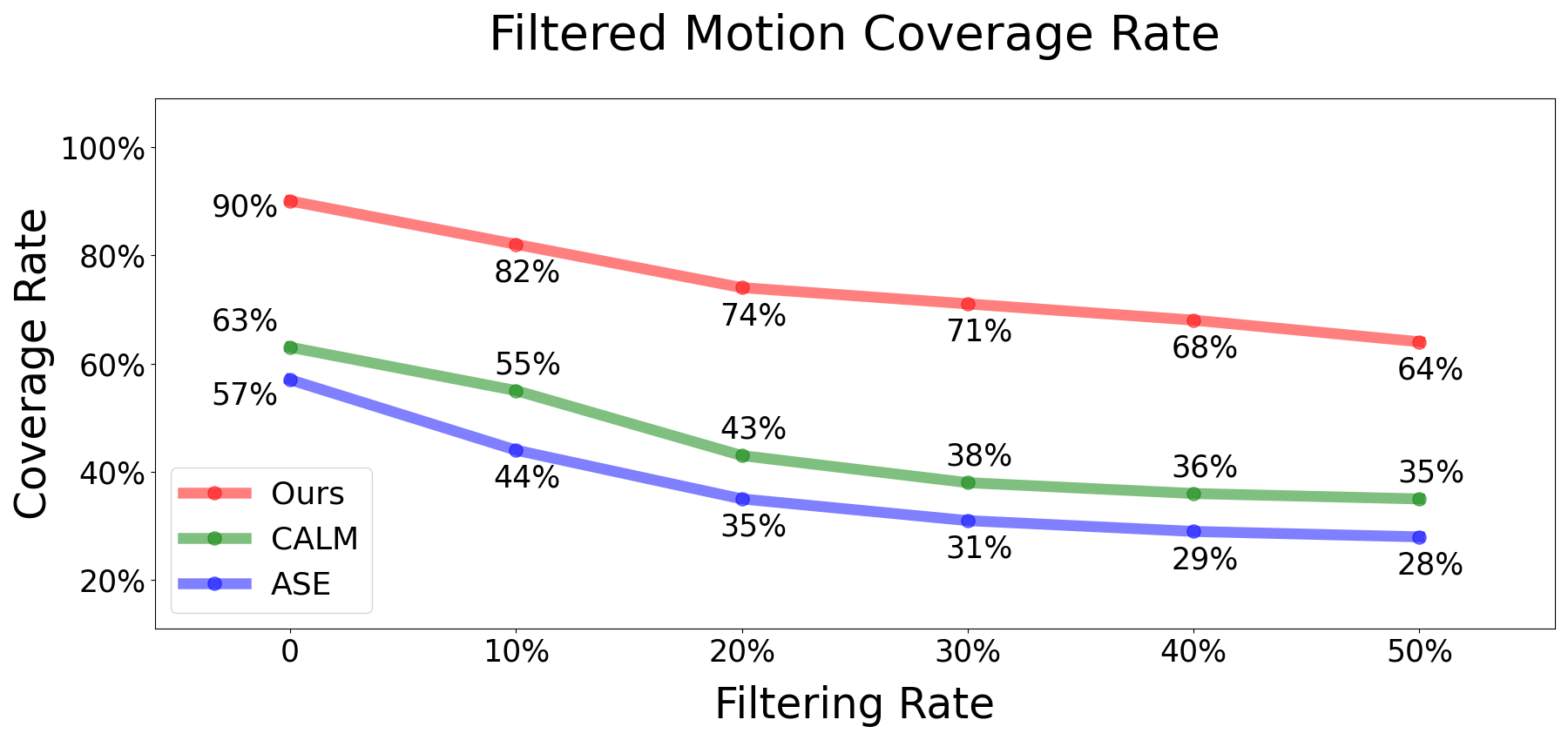}
\end{overpic}
\caption{Comparison of the motion coverage on the Composite Skills dataset with 778 clips of 265 motion skills. The coverage rate of ASE and CALM falls dramatically with an increasing filtering rate, implying a serious unbalance of the coverage, whereas ours consistently produces high motion coverage rates.}
\label{fig:motion_covrage_cutoff_curve_all_mixed}
\end{figure}

\begin{figure}[!t]
\centering
\begin{overpic}[width=\linewidth]{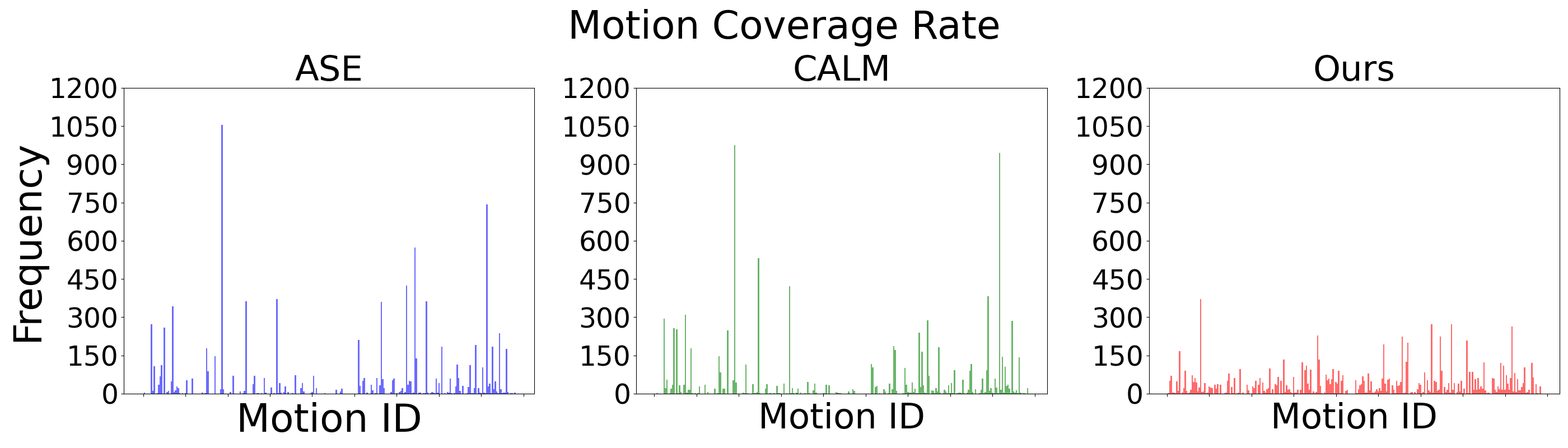}
\put(41,-3){(a) No filtering.}
\end{overpic}
  \makebox[\linewidth][c]{}\\
  \makebox[\linewidth][c]{}\\
  \begin{overpic}[width=\linewidth]{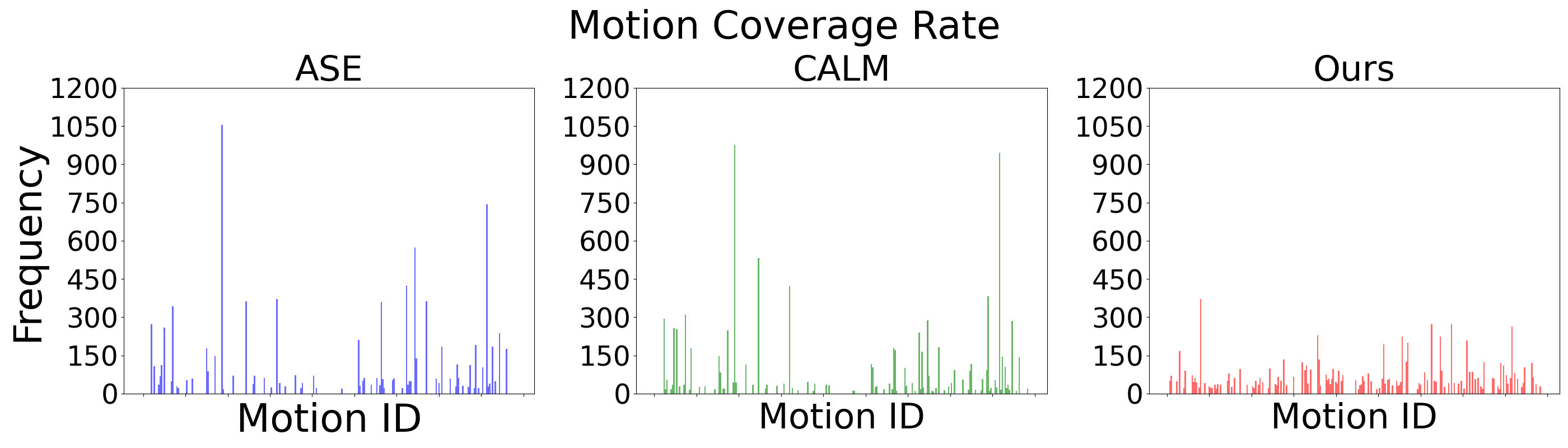}
  \put(33,-3){(b) Filtering rate $= 50\%$.}
\end{overpic}
  \makebox[\linewidth][c]{}\\
\caption{Frequencies at which the low-level policy produces motions that match all 265 motion skills in the Composite Skills dataset. We show distributions produced by the filtering rate of 0\% and 50\%. Compared to ASE and CALM, our method produces diverse motions that much more evenly cover all reference clips.}
\label{fig:motion_covrage_bar_all_mixed}
\vspace{-3mm}
\end{figure}

Figure~\ref{fig:motion_covrage_cutoff_curve_all_mixed} presents the motion coverage rate calculated using different \filteringrates.
We can see ASE's coverage dramatically drops to $44\%$, $35\%$, and $28\%$ at the filtering rate of $10\%$, $20\%$, and $50\%$ \filteringrate, respectively. The coverage rate of CALM drops from $55\%$ to $43\%$, and $35\%$ at the filtering rate of $10\%$, $20\%$, and $50\%$ \filteringrate. Meanwhile, we observed that CALM required more training steps to achieve a reasonable motion coverage rate.
This implies a serious imbalance and instability of the motion coverage, i.e., many motion clips are matched only a few times, possibly due to stochastic factors existing in the randomly generated trajectories, as evidenced by Figure~\ref{fig:motion_covrage_bar_all_mixed}, that records the frequencies at which $\policy$ produces trajectories that
matched each motion clip in the dataset across $10,000$ trajectories.

\subsection{Skill Label}
\label{ap:skill_acquisition}

\subsubsection{Skill Label Acquisition}
\begin{figure*}
\centering
\begin{overpic}[width=\linewidth]{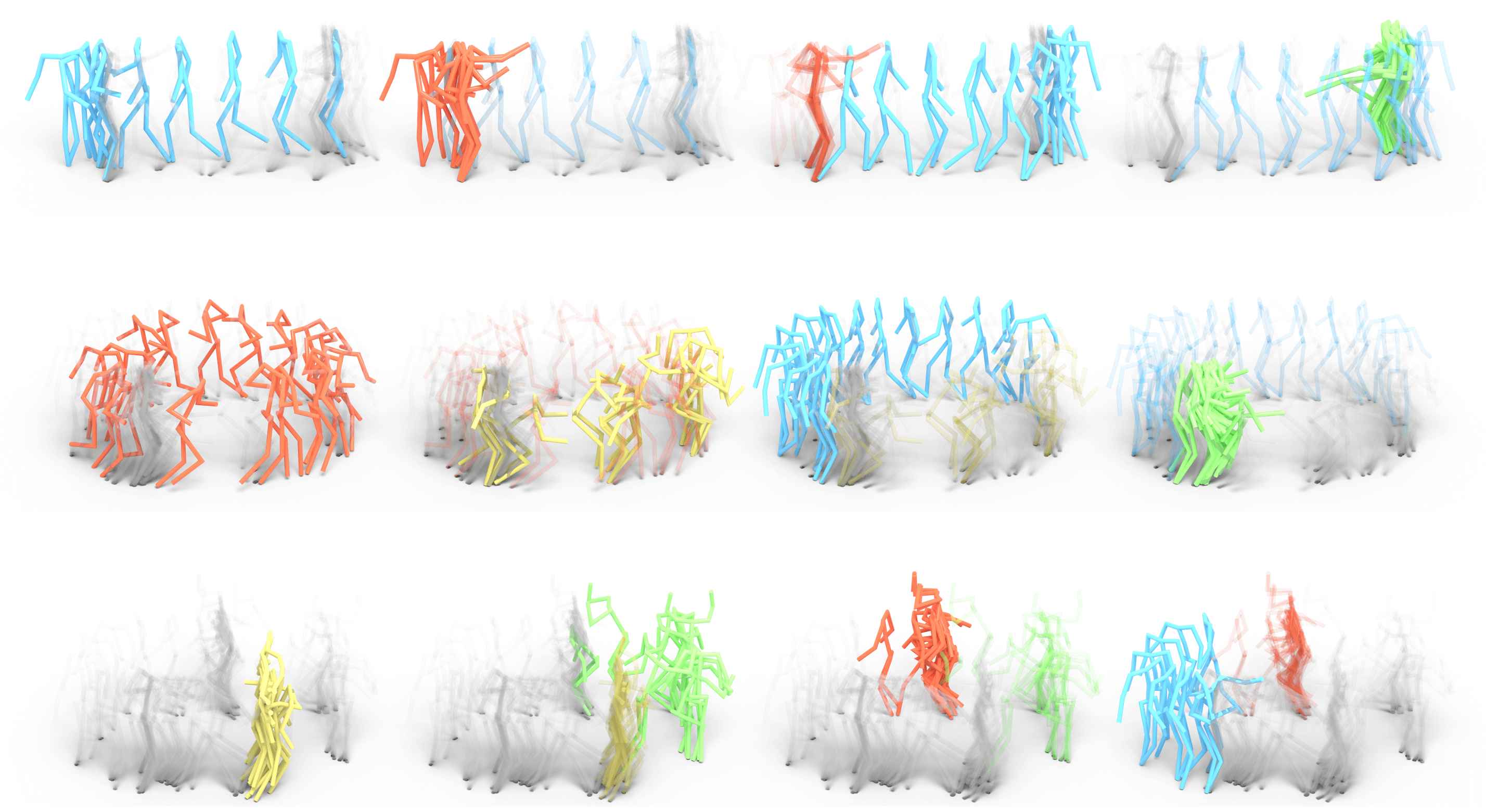}
\put(11,38){(a) Walk}
\put(34,38){(b) Punch}
\put(59,38){(c) Walk}
\put(84,38){(d) Kick}

\put(11.5,18){(a) Run}
\put(35,18){(b) Hop}
\put(59,18){(c) Walk}
\put(84,18){(d) Kick}

\put(10,-2){(a) Jack Jump}
\put(35,-2){(b) Jump}
\put(59,-2){(c) Hop}
\put(84,-2){(d) Walk}
\end{overpic}
\makebox[0.5\linewidth][c]{}
\caption{Motion Segmentation results on unstructured motion clips. Each row represents one unstructured motion clip from the CMU Mocap dataset~\cite{CMU}. We highlight each segmented skill in a specific color.}
\label{fig:motion_segmentation}
\vspace{8mm}
\end{figure*}

\begin{figure*}
\centering
\begin{overpic}[width=\linewidth]{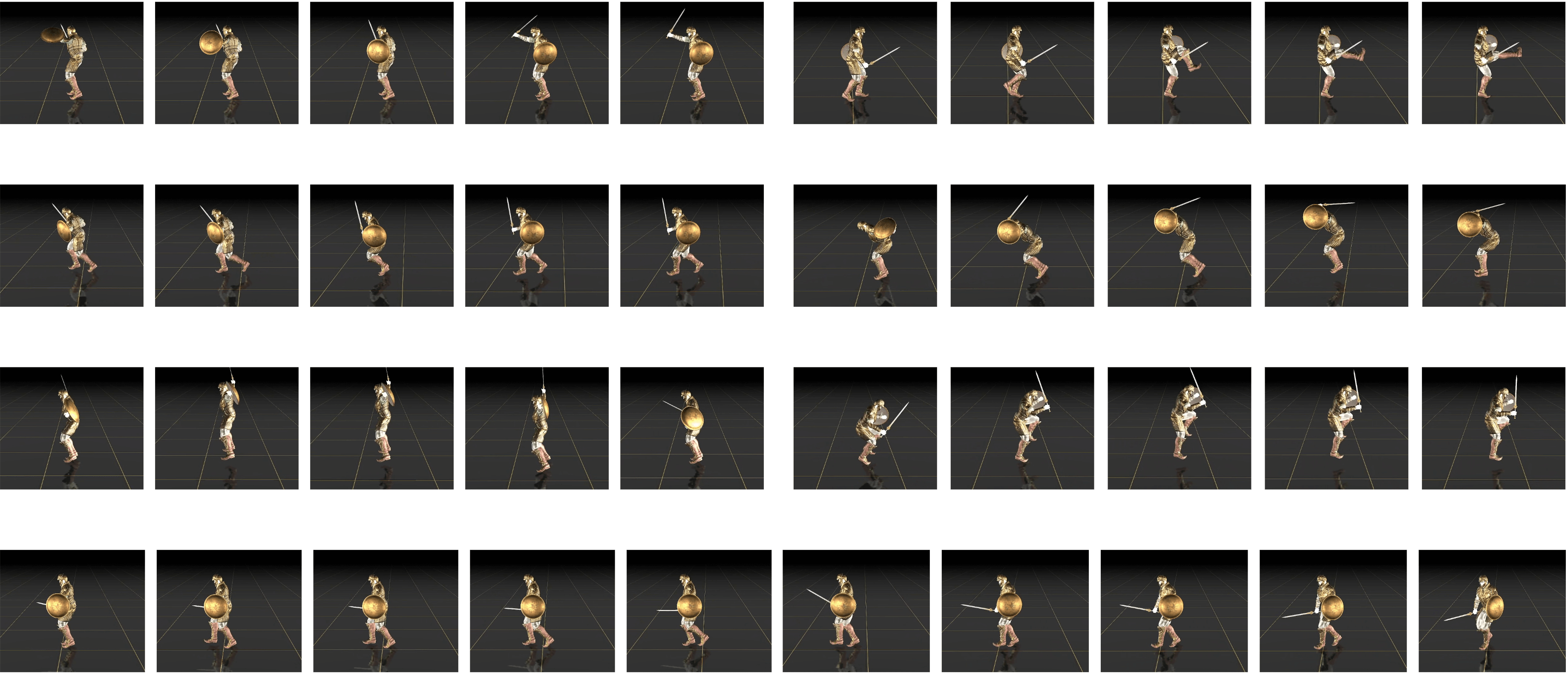}
\put(22,33.5){Punch}
\put(74,33.5){Kick}
\put(23,21.5){Run}
\put(74,21.5){Jump}
\put(21,10){Jack Jump}
\put(74,10){Hop}
\put(48,-2){Walk}
\end{overpic}
\makebox[0.5\linewidth][c]{}
\caption{Various skills performed by \name under different skill conditions based on the motion segmentation results.}
\label{fig:motion_segmentation_skills}
\end{figure*}
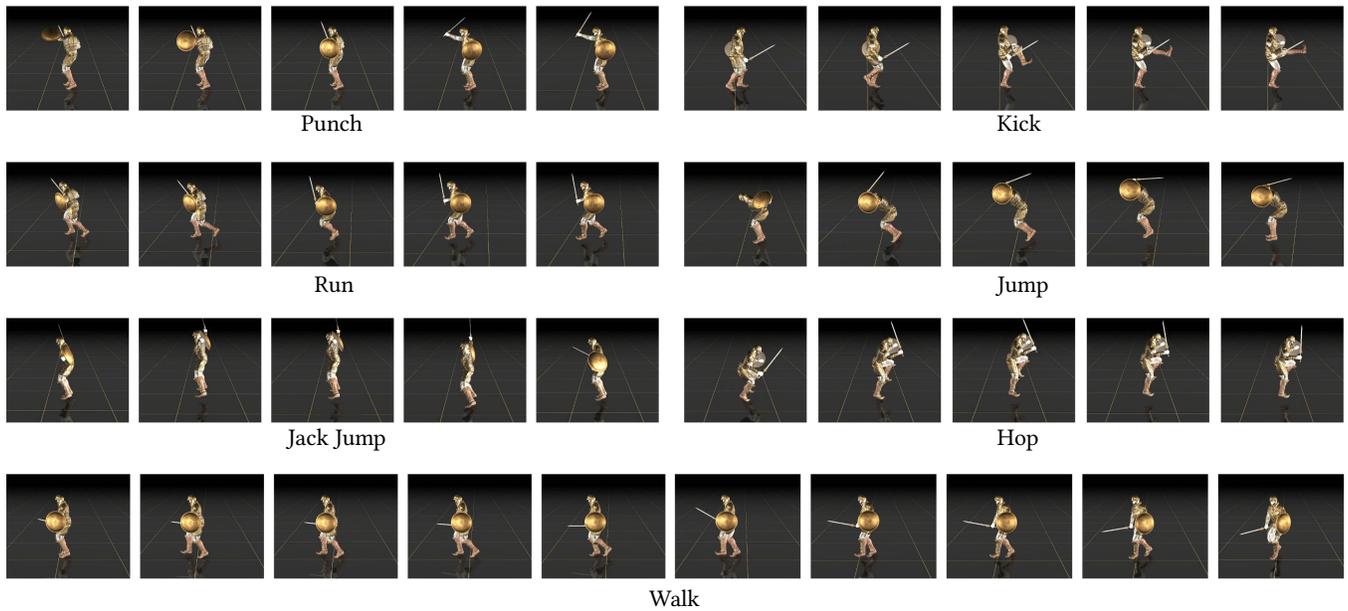

We investigate the performance of \name on unorganized mocap data. The key to processing unorganized motion datasets is label acquisition. We demonstrate that an action recognition network~\cite{duan2022pyskl} pretrained on NTURGB$+$D 120 dataset~\cite{liu2020ntu} can produce reliable skill labels. We test with raw unstructured motion clips from CMU Mocap dataset~\cite{CMU} by finetuning the network on the aforementioned Composite Skills dataset, where we set the learning rate to be $0.1$ with momentum being $0.9$ and weight decay being $5\times10^{-4}$. The model is trained on a single A100 GPU. We set the batch size to be $128$ and train for $80$ epochs. We randomly divided the data set into the training set and test set by 9$:$1, and the top-1 accuracy on the test set was $76.9\%$. As illustrated in Figure~\ref{fig:motion_segmentation}, \name effectively embeds the segmented skills. Note that the data in Figure~\ref{fig:motion_segmentation} is taken from the test set. As demonstrated in Figure~\ref{fig:motion_segmentation_skills}, our approach successfully captured the reference skills corresponding to each skill condition from the original unstructured motion clips.

In fact, using the skill label as a condition brings a lot of flexibility, which makes our method compatible with various input sources. For instance, methods that directly regress the skeleton rotation~\cite{shi2020motionet} or joint rotation~\cite{li2021hybrik, dou2022tore, cai2022humman, wang2023zolly, li2023niki,wan2021encoder} of the parametric model, e.g., SMPL~\cite{loper2015smpl}, from the video, can be directly applied to our labeling.  Moreover, skill labels are typically aligned with the language of skill descriptions, which could open the door for language-guide motion control.

\subsubsection{Random Labeling}
\label{ap:random_labeling}
Regarding the influence of labeling, we investigate the performance of \name at different levels of randomness where we train \name on Sword\&Shield dataset containing $87$ motions with fewer labels:  that are \textit{randomly} assigned to each clip. Due to the randomness of labeling, samples under each skill label are more heterogeneous than those in our default setting. The result can be found in Table~\ref{ap_tab:randomness}. From Table~\ref{ap_tab:randomness}, we could see that by randomly grouping the skills with assigned labels for skill embedding, the motion coverage may improve to a certain extent, but not significantly. Specifically, when the number of random labels is small (fewer groups), motion clips under each group are more heterogeneous. Thus, the improvement in the efficiency of skill embedding becomes limited. In fact, an efficient conditional skill embedding calls for not only skill partitions but also to ensure that the motions within each skill group are as homogeneous as possible.

\begin{table}[t!]
  \caption{ Filtered motion coverage of \name with randomly assigned labels. We report the performance at different filtering rates.}
  \label{ap_tab:randomness}
\begin{tabular}{c|cccccc}
    \toprule
\multirow{2}{*}{{\makecell{\# Random\\ Label}}} & \multicolumn{6}{c}{Filtering Rate} \\
& $ 0 $ & $10\%$ & $20\%$  & $30\%$ &  $40\%$&  $50\%$\\ 
    \midrule
16 &$ 78 $&$ 57 $&$ 45 $&$ 43 $&$ 41 $&$ 40 $\\ 
32 &$ 87 $&$ 62 $&$ 55 $&$ 52 $&$ 51 $&$ 49 $\\ 
64 &$ 90 $&$ 83 $&$ 79 $&$ 77 $&$ 77 $&$ 75 $\\ 
    \toprule
\end{tabular}
\end{table}

\subsection{More High-Level Tasks}
\label{ap:high_level_no_user}
In the following, we evaluate \name in various representative high-level tasks, such as \textit{Reach}, \textit{Steering}, \textit{Location}, and \textit{Strike}, following~\cite{ase}. For detailed task configurations and goal reward designs, we refer readers to~\cite{ase}.
Notably, in this paper, we automatically filter out motion clips whose variation of projected root trajectory on XY-plane is less than 0.3m within one episode for various downstream tasks with an emphasis on locomotion skills.


\paragraph{Hybrid Action Space.} To achieve the task-training objective, the high-level policy learns to output an optimal configuration of both the skill label $\skilllabel$ and the latent transition code $\skilllatent$, sampled from the \emph{discrete} skill set $\skillset$ and the \emph{continuous} latent space $\skilllatentspace$. Unlike previous work~\cite{ase, controlvae} with purely continuous action spaces, our high-level policy training falls into discrete-continuous action reinforcement learning~\cite{fan2019hybrid, li2021hyar}. Directly regressing the discrete skill label and latent transition code hinders learning and results in poor performance. Instead, we treat skill label prediction as a classification problem, with the high-level policy outputting a continuous probability distribution over skill labels. The skill label $\skilllabel$ is predicted as a continuous probability distribution and converted to the skill label index using Gumbel-Softmax~\cite{jang2016categorical} for the low-level controller, while latent $\skilllatent$ is drawn from a spherical Gaussian distribution.

\begin{figure*}
\centering
\begin{overpic}[width=\linewidth]{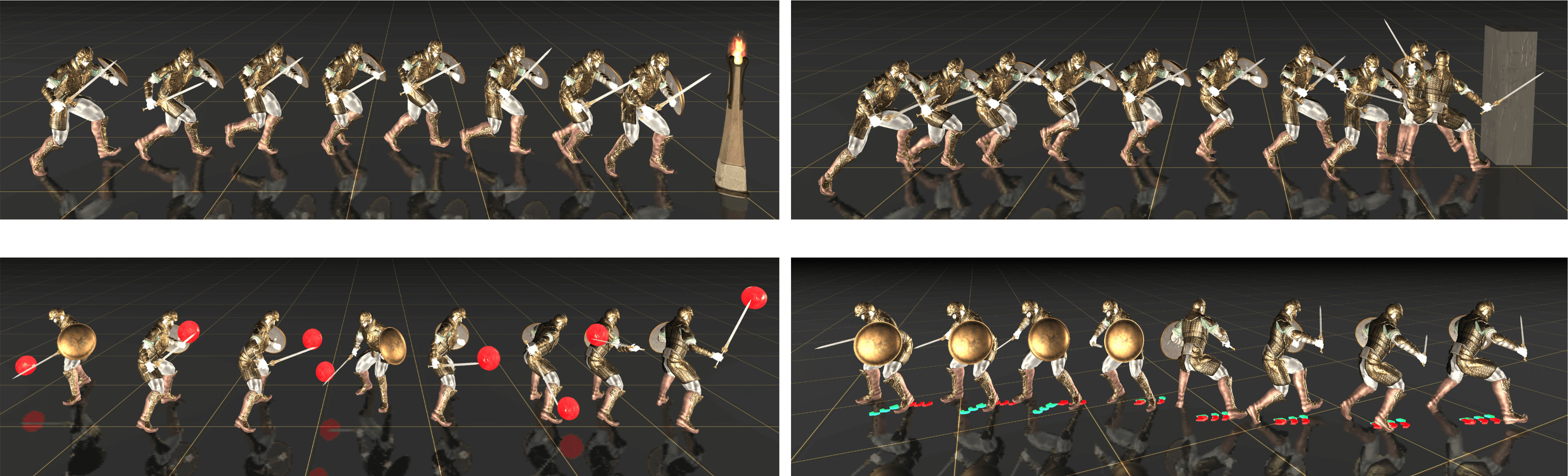}
\put(20.5,+14.7){(a) Location}
\put(71.5,+14.7){(b) Strike}
\put(21.5,-1.6){(c) Reach}
\put(71,-1.6){(d) Steering}
\end{overpic}
\caption{
A physically Simulated character can be trained to complete tasks using skills from the pre-trained low-level policy. The character successfully achieves the goals of high-level tasks while producing naturalistic behaviors. (a) Location: the character moves towards the target location, i.e., torch. 
(b) the character moves to the target object and knocks it over with its sword. 
(c) Reach: the character positions the tip of the sword at a target location. 
(d) Steering: the character moves along a target direction (red arrow) while facing a target heading direction (green arrow).}
\label{fig:high_level_motion}
\end{figure*}
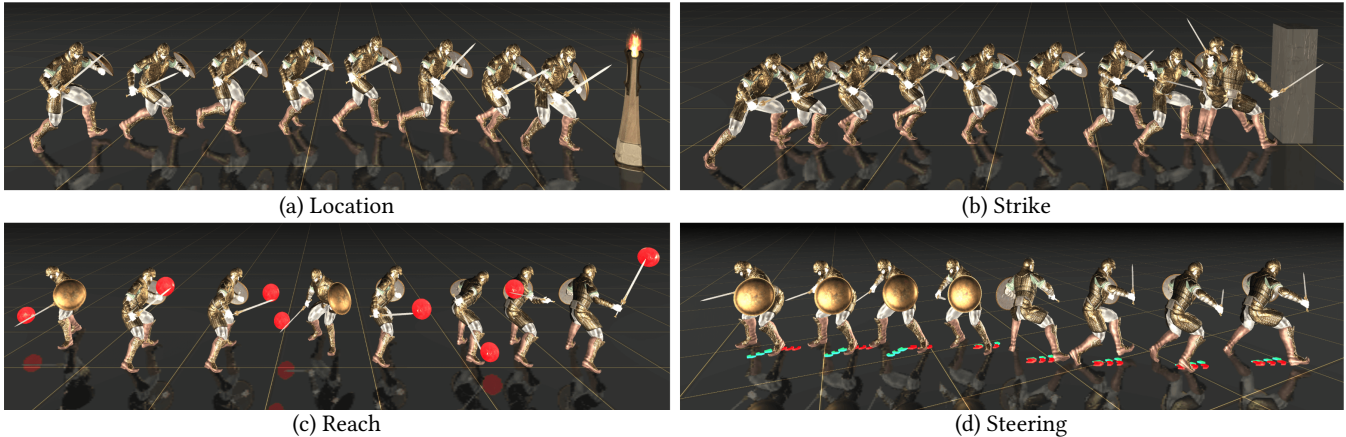

Following ASE~\cite{ase}, we also reuse the pre-trained low-level model and train a separate high-level policy for each task with no human intervention, including \emph{Location}, \emph{Strike}, \emph{Reach}, and \emph{Steering}. As depicted in Figure~\ref{fig:high_level_motion}, our model successfully accomplishes the given tasks while maintaining realistic motions, thanks to the pre-trained conditional adversarial embeddings. Specifically, when trained to strike a target object, the character seamlessly transitions from the locomotion state (e.g., running) to the sword attack skill, effectively and naturally completing the task. The character is also capable of resisting external perturbations; see more results in the supplementary video.

\subsection{Implementation Details}
\label{ap:implementation}

\subsubsection{Definition of State and Action}
\label{ap:state_and_action}

The state $\states$ for the discriminator $\Discriminator$ and encoder $\encoderq$ is defined by the set of the height of the root from the ground, rotation of the root in the local coordinate frame, linear and angular velocity in the local coordinate frame, local rotation of each joint, local velocity of each joint,  positions of hands, feet in the local coordinate frame. For low-level policy network $\policy$ and high-level policy network $\hpolicy$, the state is additionally with the position of the shield and the tip of the sword in the local coordinate frame. The action $\action$ is in $31$-D, specifying the target rotations for the PD controller at each joint. 

\subsubsection{Condition Randomization.}
At the beginning of each episode, we randomize the $(\states, \skilllabel)$ pairs independently, selecting a random reference state from the dataset and a random skill label from the set of skill labels. This approach forces the character to perform the desired skill from an arbitrary state while producing realistic transitions between various skills, as we shall demonstrate. Similar strategies have been used in~\cite{peng2018deepmimic, sharon2005synthesis, nair2018overcoming, rajeswaran2017learning}.

In the context of GAN, Randomly Initialized State (RIS) and Random Condition Signals (RCS) are applied to the policy that serves as a generator. Since the discriminator is taking paired reference motions and their condition signals, the generator~(policy) will learn to match each condition signal to the corresponding motion. Combinatorial variations brought by RIS and RCS enable the policy to be robust and insensitive to the exact boundary of labels and allow for natural transitions between various skills, as shown in the video. We found that a bigger and more diverse dataset, i.e., the Composite Skills dataset, will not reduce our performance.

\subsubsection{Network Structure}
\label{ap:network_structure}
 We implement the policy networks $\policy$ and $\hpolicy$, value function $V$, encoder $\encoderq$ and discriminator $\Discriminator$ using MLP.  Specifically, the low-level policy $\policy$ is modeled by a neural network that maps a state $\states$ and latent $\skilllatent$ and skill label $\skilllabel$ to a Gaussian distribution over actions with a fixed diagonal covariance matrix. The network is implemented by the MLP with ReLU units and hidden layers containing $[1024, 1024, 512]$. Note that the discrete skill label $\skilllabel$ is embedded into a $64$D space and concatenated to the input before being passed to the first layer of the network. A similar structure is used for the value function $V$ but with a single linear output unit.

Different from ASE~\cite{ase} that models the encoder and discriminator as one single network with different output heads, in our model, the encoder is modeled as one network using MLP of hidden layers containing $[1024, 1024, 512]$ with ReLU units while the discriminator is implemented in another network using MLP of hidden layers containing $[1024, 512]$ with ReLU units for discrimination. To condition the discriminator $\Discriminator$, we increase the dimension of the output layer to be the number of skills and then the feature specified by $\skilllabel$ is used for discrimination. The conditional encoder is implemented similarly to $\policy$, where we embed the skill label and concatenate it to the input.

The high-level policy $\hpolicy$ is implemented with 2 hidden layers with [1024, 512] units, followed by a linear layer to predict the mean of latent code and skill label. Then the discrete skill label is obtained by taking the class with the highest probability.

\subsection{Training Details}
\label{ap:training_details}
For each skill label $\skilllabel$, we embed it into a $64$D latent space. And the latent space for $\skilllatent$ is $16$D. The random probability for Element-wise Feature Masking is set to be $\maskingrate =20\%$ while the update rate $\forgetrate$ is set to be $20\%$ in Focal Skill Sampling. The other hyperparameter settings for low-level policy and high-level policy, as well as corresponding value functions, are the same as ASE~\cite{ase}. Following ASE, we also randomize the sequence of latent codes during low-level policy training. Each latent is fixed for between $1$ and $150$ time steps before the update. The buffer length for the discriminator and encoder is set to be $20$. During task training, the frequency of high-level policy and low-level policy are $6$Hz and $30$Hz, respectively. The skill label and latent produced by the high-level policy are repeated for $5$ steps for the low-level policy. 

We integrate reference motion from the CMU dataset into the conditional discriminator using adaptive masking for sword and shield components. Specifically, as the skill label serves as a condition when obtaining skill labels from the Sword\&Shield dataset, we consider key points, including those for the sword and shield, during the discriminator process. For skills derived from the CMU Mocap dataset, we apply masking to the sword and shield key points for both reference motion (real samples) and physically simulated motion (fake samples). This approach eliminates the need for the simulated character's sword and shield positions to align with the reference motion, as these positions may be unreliable during the retargeting process from the CMU Mocap dataset.

Furthermore, we employ curriculum learning for efficient training. At the beginning of the training, we disallow body-ground contact for the first $15000$ epochs to accelerate the training process, then permit it in subsequent epochs to learn those hard-to-learned movements. 








\subsection{Failure Cases}
\label{ap:limitations}
In this section, we show failure cases, e.g., characters only launch idling motion when conditioned on a skilled label that is not embedded well. Details can be found in our supplementary video. We also observed that there is a trend of coverage improving these skills in \name as training proceeds.

\end{document}